\documentclass[aps,nofootinbib]{revtex4}

\usepackage{graphicx}
\usepackage{graphics}
\usepackage{amssymb}
\usepackage{amsmath}
\usepackage{bm}

\newcommand{\half}{\textstyle{\frac{1}{2}}}

\newcommand{\be}{\begin{equation}}
\newcommand{\ee}{\end{equation}}
\newcommand{\bea}{\begin{eqnarray}}
\newcommand{\eea}{\end{eqnarray}}
\newcommand{\beal}{\begin{align}}
\newcommand{\eal}{\end{align}}
\newcommand{\bespl}{\begin{split}}
\newcommand{\espl}{\end{split}}
\newcommand{\nn}{\nonumber}
\newcommand{\nslash}{\kern 0.2 em n\kern -0.50em /}
\newcommand{\kslash}{\kern 0.2 em k\kern -0.45em /}
\newcommand{\pslash}{\kern 0.2 em p\kern -0.50em /}
\newcommand{\Sslash}{\kern 0.2 em S\kern -0.50em /}
\newcommand{\Pslash}{\kern 0.2 em P\kern -0.50em /}
\newcommand{\Rslash}{\kern 0.2 em R\kern -0.50em /}
\newcommand{\open}{{<\kern -0.3 em{\scriptscriptstyle )}}}
\newcommand{\sT}{\scriptstyle T}

\begin{document}
\title{Nucleon Spin Structure with hadronic collisions at COMPASS}

\author{\underline{Marco Radici}\footnote{talk given at the "International Workshop on 
Structure and Spectroscopy", Freiburg, March 19-21, 2007.}}
\email{marco.radici@pv.infn.it}
\affiliation{Dipartimento di Fisica Nucleare e Teorica, Universit\`{a} di 
Pavia, and\\
Istituto Nazionale di Fisica Nucleare, Sezione di Pavia, I-27100 Pavia, Italy}

\author{Francesco Conti}
\email{francesco.conti@pv.infn.it}
\affiliation{Dipartimento di Fisica Nucleare e Teorica, Universit\`{a} di 
Pavia, and\\
Istituto Nazionale di Fisica Nucleare, Sezione di Pavia, I-27100 Pavia, Italy}

\author{Alessandro Bacchetta}
\email{alessandro.bacchetta@desy.de}
\affiliation{Theory Group, Deutsches Elektronen-Synchroton DESY, \\
D-22603 Hamburg, Germany}

\author{Andrea Bianconi}
\email{andrea.bianconi@bs.infn.it}
\affiliation{Dipartimento di Chimica e Fisica per l'Ingegneria e per i 
Materiali, Universit\`a di Brescia, I-25123 Brescia, Italy, and\\
Istituto Nazionale di Fisica Nucleare, Sezione di Pavia, I-27100 Pavia, Italy}

\begin{abstract}
In order to illustrate the capabilities of COMPASS using a 
hadronic beam, I review some of the azimuthal asymmetries in 
hadronic collisions, that allow for the extraction of transversity, 
Sivers and Boer-Mulders functions, necessary to explore the 
partonic spin structure of the nucleon. I also report on some 
Monte Carlo simulations of such asymmetries for the production 
of Drell-Yan lepton pairs from the collision of high-energy pions 
on a transversely polarized proton target. 
\end{abstract}

\maketitle

\section{Introduction}
\label{sec:intro}

The recent measurement of Single-Spin Asymmetries (SSA) in semi-inclusive $l
p^\uparrow \to l'\pi X$ Deep-Inelastic Scattering (SIDIS) on transversely
polarized hadronic 
targets~\cite{Airapetian:2004tw,Diefenthaler:2005gx,Avakian:2005ps,Alexakhin:2005iw}, 
has renewed the interest in the problem of describing the spin structure of 
hadrons within Quantum Chromo-Dynamics (QCD)~\cite{cerncourier}. Experimental 
evidence of large SSA in hadron-hadron collisions was well known since many 
years~\cite{Bunce:1976yb,Adams:1991cs}, but it has never been consistently 
explained in the context of perturbative QCD in the collinear massless 
approximation~\cite{Kane:1978nd}. Similarly, in a series of experiments on 
high-energy collisions of pions and antiprotons with various unpolarized 
nuclei~\cite{Falciano:1986wk,Guanziroli:1987rp,Conway:1989fs,Anassontzis:1987hk} 
the cross section for Drell-Yan events shows an unexpected largely asymmetric 
azimuthal distribution of the final lepton pair with respect to the production 
plane, which leads to a bad violation of the socalled Lam-Tung sum 
rule~\cite{Lam:1980uc}. Again, in the context of collinear perturbative QCD 
there are no calculations able to consistently justify this set of 
measurements~\cite{Brandenburg:1994wf,Eskola:1994py,Berger:1979du}. 

The idea of going beyond the collinear approximation opened new perspectives 
about the possibility of explaining these SSA in terms of intrinsic transverse 
motion of partons inside hadrons, and of correlations between such intrinsic 
transverse momenta and transverse spin degrees of freedom. The most popular 
examples are the Sivers~\cite{Sivers:1990cc} and the Collins~\cite{Collins:1993kk} 
effects. In the former case, an asymmetric azimuthal distribution of detected 
hadrons (with respect to the normal to the production plane) is obtained from the 
nonperturbative correlation ${\bm p}_{\sT} \times {\bm P}\cdot {\bm S}_{\sT}$, where 
${\bm p}_{\sT}$ is the intrinsic transverse momentum of an unpolarized parton 
inside a target hadron with momentum ${\bm P}$ and transverse polarization 
${\bm S}_{\sT}$. In the latter case, the asymmetry is obtained from the correlation 
${\bm k} \times {\bm P}_{h\sT} \cdot {\bm s}_{\sT}$, where a parton with momentum 
${\bm k}$ and transverse polarization ${\bm s}_{\sT}$ fragments into an unpolarized 
hadron with transverse momentum ${\bm P}_{h\sT}$. In both cases, the sizes of the 
effects are represented by new unintegrated, or Transverse-Momentum Dependent 
(TMD), partonic functions, the socalled Sivers and Collins functions, respectively. 

However, SSA data in hadronic collisions have been collected so far typically for
semi-inclusive $pp^{(\uparrow )}\to h^{(\uparrow )} X$ processes, where the
factorization proof is still under debate (see Ref.~\cite{Collins:2007nk} and 
references therein). On the contrary, the theoretical situation of the SIDIS 
measurements is more transparent. On the basis of a suitable factorization 
theorem~\cite{Ji:2004wu,Collins:2004nx}, the cross section at leading twist 
contains convolutions involving separately the Sivers and Collins functions with 
different azimuthal dependences, $\sin (\phi - \phi_S)$ and $\sin (\phi + \phi_S)$, 
respectively, where $\phi, \phi_S,$ are the azimuthal angles of the produced hadron 
and of the target polarization with respect to the axis defined by the virtual 
photon~\cite{Boer:1998nt}. According to the extracted azimuthal dependence, the 
measured SSA can then be clearly related to one effect or the 
other~\cite{Airapetian:2004tw,Diefenthaler:2005gx}.

Similarly, in the Drell-Yan process $H_1 H_2^\uparrow \to l^+ l^- X$ the cross
section displays at leading twist two terms weighted by $\sin (\phi - \phi_S)$
and $\sin (\phi + \phi_S)$, where now $\phi, \phi_S,$ are the azimuthal 
orientations of the final lepton plane and of the hadron polarization with 
respect to the reaction plane~\cite{Boer:1999mm}. Adopting the notations recommended 
in Ref.~\cite{Bacchetta:2004jz}, the first one involves the convolution of the 
Sivers function $f_{1\sT}^\perp$ with the standard unpolarized parton distribution 
$f_1$. The second one involves the yet poorly known transversity distribution $h_1$ 
and the Boer-Mulders function $h_1^\perp$, a TMD distribution which is most likely 
responsible for the violation of the previously mentioned Lam-Tung sum 
rule~\cite{Boer:1999mm}. Hence, a simultaneous measurement of unpolarized and 
single-polarized Drell-Yan cross sections would allow to extract all the above 
partonic densities from data~\cite{Bianconi:2004wu,Bianconi:2005px}. Both $h_1$ 
and $h_1^\perp$ describe the distribution of transversely polarized partons; but 
the former applies to transversely polarized parent hadrons, while the latter to 
unpolarized ones. On an equal footing, $f_{1\sT}^\perp$ and $f_1$ describe 
distributions of unpolarized partons. The correlation between ${\bm p}_{\sT}$ and 
${\bm S}_{\sT}$ inside $f_{1\sT}^\perp$ is intuitively possible only for a 
nonvanishing orbital angular momentum of partons. Then, extraction of Sivers 
function from SIDIS and Drell-Yan data would allow to study the orbital motion 
and the spatial distribution of hidden confined partons~\cite{Burkardt:2003je}, 
as well as to test its predicted sign change from SIDIS to Drell-Yan, a peculiar 
universality feature deduced on very general arguments that holds also for 
$h_1^\perp$~\cite{Collins:2002kn}. 
On the same footing, the prediction that the first moment of transversity, the 
nucleon tensor charge, is much larger than its helicity, as it emerges from 
lattice studies~\cite{Aoki:1996pi,Gockeler:2006zu}, represents a basic test of 
QCD in the nonperturbative domain.

In a series of 
papers~\cite{Bianconi:2004wu,Bianconi:2005bd,Bianconi:2005yj,Bianconi:2006hc}, 
numerical simulations of SSA in Drell-Yan processes were performed using 
transversely polarized proton targets, and (anti)proton and pion beams, in 
different kinematic conditions corresponding to the setup of GSI, RHIC, and 
COMPASS. After briefly reviewing the formalism in Sec.~\ref{sec:formula} and the 
strategy for Monte Carlo simulation in Sec.~\ref{sec:mc}, in Sec.~\ref{sec:dyssa} 
we will reconsider the $\sin (\phi \pm \phi_S)$ SSA for the 
$\pi^\pm p^\uparrow \to \mu^+ \mu^- X$ process using the pion beam at COMPASS, 
in order to verify if the foreseen statistical accuracy is enough to test the 
predicted sign change of $f_{1\sT}^\perp$, and to extract information 
on $h_1^\perp$.

In the SIDIS production of pions on transversely polarized targets, the above 
mentioned Collins effect represents the most popular technique to extract the 
transversity $h_1$ from a SSA measurement. However, it requires the cross section 
to depend explicitly upon ${\bm P}_{h\sT}$, the transverse momentum of the detected 
pion with respect to the photon axis~\cite{Boer:1998nt}. This fact brings in several 
complications, including the possible overlap of the Collins effect with other 
competing mechanisms and more complicated factorization proofs and evolution 
equations~\cite{Collins:2004nx,Ji:2004wu}. Semi-inclusive production of two 
hadrons~\cite{Collins:1994kq,Jaffe:1998hf} offers an alternative framework, where 
the chiral-odd partner of transversity is represented by the Dihadron Fragmentation 
Function (DiFF) $H_1^{\open}$~\cite{Radici:2001na}, which relates the transverse 
spin of the quark to the azimuthal orientation of the two-hadron plane.
This function is at present unknown. Very recently, the HERMES collaboration has
reported measurements of the asymmetry containing the product
$h_1\ H_1^{\open}$~\cite{vanderNat:2005yf}. The COMPASS collaboration has
also presented analogous preliminary results~\cite{Joosten:2005}. In the meanwhile, 
the BELLE collaboration is planning to measure $H_1^{\open}$ in the near 
future~\cite{Abe:2005zx}. A spectator model calculation of leading-twist DiFF has 
been built by tuning the parameters on the output of pion pair distributions from 
PYTHIA, adapted to the HERMES kinematics~\cite{Bacchetta:2006un}. Evolution equations 
for DiFF have been also studied, including the full dependence upon the pair 
invariant mass~\cite{Ceccopieri:2007ip}. 

In Sec.~\ref{sec:ppDiFF}, we will reconsider the proposal formulated for the first 
time in Ref.~\cite{Bacchetta:2004it}, namely to use hadronic collisions on a 
transversely polarized proton target and inclusively detect one (or two) pairs of 
pions. By combining unpolarized and polarized cross sections, in principle it is 
possible to "self-determine" all unknown DiFF in one experiment and to extract 
the transversity $h_1$ in a clean way. We think that the possibility should be 
explored to perform such a program at COMPASS using a pion beam.

\begin{figure}[h]
\centering
\includegraphics[width=7cm]{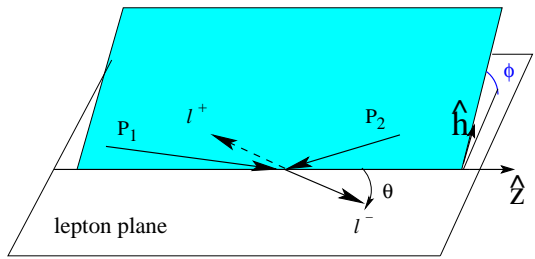}
\caption{The Collins-Soper frame.}
\label{fig:dyframe}
\end{figure}


\section{The formalism for the Drell-Yan cross section}
\label{sec:formula}

In the following, we will just sketch the kinematics and the necessary formulae 
for describing the $H_1 H_2^\uparrow \to l^+ l^- X$ cross section at leading twist. 
We forward the interested reader to Ref.~\cite{Bianconi:2006hc} for the technical 
details.

In the collision of two hadrons with momentum $P_1, P_2$, and masses $M_1, M_2$, 
the center-of-mass (c.m.) square energy $s=(P_1+P_2)^2$ is available to produce two 
leptons with momenta $k_1, k_2$, and invariant mass $q^2 \equiv M^2 = (k_1 + k_2)^2$. 
In the limit  $M^2,s \rightarrow \infty$, while keeping the ratio 
$0\leq \tau = M^2/s \leq 1$ limited, a factorization theorem can be 
proven~\cite{Collins:1984kg} ensuring that the elementary mechanism proceeds through 
the annihilation of a parton and an antiparton into a virtual photon with time-like 
momentum $q^2$. The partons are characterized by fractional longitudinal momenta 
$x_1, x_2$, and spatial trasverse components ${\bm p}_{i\sT}$ with respect to the 
direction defined by ${\bm P}_i \, (i=1,2)$, respectively. Momentum conservation implies 
${\bm q}_{\sT} = {\bm p}_{1\sT} + {\bm p}_{2\sT}$. If ${\bm q}_{\sT} \neq 0$ the 
annihilation direction is not known. Hence, it is convenient to select the 
socalled Collins-Soper frame~\cite{Collins:1977iv} described in 
Fig.~\ref{fig:dyframe}. The final lepton pair is detected in the solid angle 
$(\theta, \phi )$, where, in particular, $\phi$ (and all other azimuthal angles) 
is measured in a plane perpendicular to the indicated lepton plane but containing 
$\hat{\bm h} = {\bm q}_{\sT} / |{\bm q}_{\sT}|$.

The full expression of the leading-twist differential cross section can be written 
as~\cite{Boer:1999mm}
\bea
\frac{d\sigma}{d\Omega dx_1 dx_2 d{\bm q}_{\sT}} &= &
\frac{d\sigma^o}{d\Omega dx_1 dx_2 d{\bm q}_{\sT}} + 
\frac{d\Delta \sigma^\uparrow}{d\Omega dx_1 dx_2 d{\bm q}_{\sT}}  \nn \\
&= &\frac{\alpha^2}{3Q^2}\,\sum_q\,e_q^2\,\Bigg\{ A(y) \, 
{\cal F}\left[ f_1^q(H_1)\, f_1^q (H_2) \right] \nn \\
& &\mbox{\hspace{2cm}} + B(y) \, \cos 2\phi \, 
{\cal F}\left[ \left( 2 \hat{\bm h}\cdot {\bm p}_{1\sT} \, \hat{\bm h} \cdot 
{\bm p}_{2\sT} - {\bm p}_{1\sT} \cdot {\bm p}_{2\sT} \right) \, 
\frac{h_1^{\perp\,q}(H_1)\,h_1^{\perp\,q}(H_2)}{M_1\,M_2}\,\right] \Bigg\} \nn \\
& &+ \frac{\alpha^2}{3Q^2}\,|{\bm S}_{2\sT}|\,\sum_q\,e_q^2\,\Bigg\{ 
A(y) \, \sin (\phi - \phi_{S_2})\, {\cal F}\left[ \hat{\bm h}\cdot 
{\bm p}_{2\sT} \,\frac{f_1^q(H_1) \, f_{1\sT}^{\perp\,q}(H_2^\uparrow)}{M_2}\right] 
\nn \\
& &\mbox{\hspace{3cm}} - B(y) \, \sin 
(\phi + \phi_{S_2})\, {\cal F}\left[ \hat{\bm h}\cdot {\bm p}_{1\sT} \,
\frac{h_1^{\perp\,q}(H_1) \, h_1^q(H_2^\uparrow)}{M_1}\right] \nn \\
& &\mbox{\hspace{3cm}}  - B(y) \, \sin (3\phi - \phi_{S_2})\, {\cal F}\left[ 
\left( 4 \hat{\bm h}\cdot {\bm p}_{1\sT} \, (\hat{\bm h} \cdot 
{\bm p}_{2\sT})^2 - 2 \hat{\bm h} \cdot {\bm p}_{2\sT} \, {\bm p}_{1\sT} \cdot 
{\bm p}_{2\sT} - \hat{\bm h}\cdot {\bm p}_{1\sT} \, {\bm p}_{2\sT}^2 \right) 
\right. \nn \\
& &\mbox{\hspace{7cm}} \left. \times
\frac{h_1^{\perp\,q}(H_1) \, h_{1\sT}^{\perp\,q}(H_2^\uparrow)}{2 M_1\,M_2^2}\,\right]
\, \Bigg\} \; ,
\label{eq:xsect}
\eea
where $\alpha$ is the fine structure constant, $d\Omega = \sin \theta d\theta d\phi$, 
$e_q$ is the charge of the parton with flavor $q$, $\phi_{S_i}$ is
the azimuthal angle of the transverse spin of hadron $i$, and 
\begin{align}
A(y) = \left( \frac{1}{2} - y + y^2 \right) \, \stackrel{\mbox{cm}}{=}\, 
\frac{1}{4}\left( 1 + \cos^2 \theta \right) &\mbox{\hspace{2cm}} 
B(y) = y (1-y) \, \stackrel{\mbox{cm}}{=}\,\frac{1}{4}\, \sin^2 \theta \; . 
\label{eq:lepton}
\end{align}

The convolutions are defined as 
\be
{\cal F} \left[ DF_1^q(H_1) \, DF_2^q(H_2^{(\uparrow )}) \right] \equiv \int 
d{\bm p}_{1\sT} d{\bm p}_{2\sT}\, \delta \left( {\bm p}_{1\sT} + {\bm p}_{2\sT} - 
{\bm q}_{\sT} \right) \, \left[ DF_1(x_1,{\bm p}_{1\sT}; \bar{q}/H_1)\, 
DF_2(x_2,{\bm p}_{2\sT}; q/H_2^{(\uparrow )} ) + (q\leftrightarrow \bar{q}) \right] 
\; .
\label{eq:convol}
\ee

Parton distribution functions can be obtained from the color-gauge invariant 
quark-quark correlator
\be
\Phi (x,S) = \int \frac{d\xi^-}{2\pi}\,e^{ixP^+\xi^-}\,\langle P,S|
\bar{\psi}(0)\, U_{[0,\xi^-]}\, \psi(\xi)|P,S\rangle \Bigg\vert_{\xi^+ =
{\bm \xi}_{\sT} = 0} \; , 
\label{eq:phi}
\ee
where

\begin{figure}[h]
\centering
\includegraphics[width=7cm]{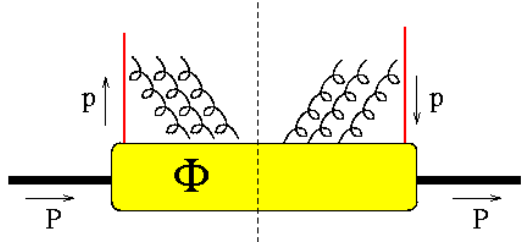}
\caption{A possible representation of the color-gauge invariant quark-quark
correlator.}
\label{fig:link}
\end{figure}

\be
U_{[0,\xi]} = {\cal P}\, e^{-ig \int_0^\xi dw \cdot A(w)}
\label{eq:link}
\ee
is the socalled gauge link operator, connecting the two different space-time points 
$0,\xi,$ by all the possible ordered paths (${\cal P}[..]$) followed by the gluon 
field $A$, which couples to the quark field $\psi$ through the constant $g$. 
Only $A^+$ gluons appear at any power in the expansion of the exponential, leading 
to an intuitive representation of the gauge link operator as a sort of residual 
interaction between the parton and the residual hadron (see Fig.~\ref{fig:link}). 
When also including the ${\bm p}_{\sT}$ dependence in Eq.~\ref{eq:phi}, the 
integration path of the gauge link is more complicated and involves ${\bm A}_{\sT}$ 
gluons at light-cone infinity~\cite{Bomhof:2004aw}. For the specific case of interest 
of hadrons with transverse polarization ${\bm S}_{\sT}$, the quark-quark correlator can 
be parametrized as 
\bea
\Phi (x, {\bm p}_{\sT},S_{\sT}) &= &\frac{1}{2P^+}\, 
\left\{  f_1 (x,{\bm p}_{\sT}) \, \Pslash + f_{1\sT}^\perp (x,{\bm p}_{\sT}) \,
\epsilon_{\mu \nu \rho \sigma}\, \gamma^\mu\, \frac{P^\nu p^\rho S_{\sT}^\sigma}{M_H} 
+ h_1^\perp (x,{\bm p}_{\sT}) \, \sigma_{\mu \nu} \, 
\frac{p^\mu P^\nu}{M_H} \right. \nonumber \\
& &\left. - h_1 (x,{\bm p}_{\sT}) \, i \sigma_{\mu \nu}\, 
S_{\sT}^\mu P^\nu \, \gamma_5 - h_{1\sT}^\perp (x,{\bm p}_{\sT})\, 
\left( \frac{{\bm p}_{\sT}\cdot {\bm S}_{\sT}}{M_H}\,
\frac{p^\mu P^\nu}{M_H} - \frac{{\bm p}_{\sT}^2}{M_H^2}\,S_{\sT}^\mu P^\nu \right) \, 
i\sigma_{\mu \nu} \, \gamma_5 \right\} \; ,
\label{eq:phiT}
\eea
where $M_H$ is the parent hadron mass. 

The TMD functions of Eq.~(\ref{eq:xsect}) can thus be obtained by suitable projections
\be
\Phi^{\left[ \Gamma \right]} (x,{\bm p}_{\sT}, {\bm S}_{\sT}) = \frac{1}{2} \int dp^- \, 
\mathrm{Tr} \left[ \Gamma\,\Phi (p,P,{\bm S}_{\sT} ) \right] \Bigg\vert_{p^+=xP^+} \; .
\label{eq:phiGamma}
\ee
In fact, following Ref.~\cite{Bacchetta:2004jz} we can define the number density 
$f_{q s_{T}/H S_{T}}$ of partons with flavor $q$ and transverse 
polarization $s_{\sT}$ in a hadron $H$ with transverse polarization $S_{\sT}$:
\bea
f_{q s_{T} / H S_{T}} &= &\frac{1}{2}\, 
\left( \Phi^{[ \nslash_{\scriptscriptstyle -} / 2 ]} + 
\Phi^{[ i\sigma_{\mu \nu}\, n_{\scriptscriptstyle -}^\mu s^\nu \gamma_5 
/2 ]}  \right) \nonumber \\
&= &\frac{1}{4}\,\left\{ f_1^q(x,{\bm p}_{\sT}) - f_{1\sT}^{\perp\, q}(x,{\bm p}_{\sT}) \, 
\frac{\hat{{\bm P}}\times {\bm p}_{\sT}\cdot {\bm S}_{\sT}}{M_H} - 
h_1^{\perp\, q}(x,{\bm p}_{\sT})\, \frac{\hat{{\bm P}}\times {\bm p}_{\sT}\cdot 
{\bm s}_{\sT}}{M_H} \right. \nonumber \\
& &\left. + h_1^q(x,{\bm p}_{\sT}) \,{\bm S}_{\sT}\cdot {\bm s}_{\sT} + 
h_{1\sT}^{\perp\, q}(x,{\bm p}_{\sT})\, \left( \frac{{\bm p}_{\sT}\cdot 
{\bm S}_{\sT}}{M_H}\, \frac{{\bm p}_{\sT}\cdot {\bm s}_{\sT}}{M_H}\, - 
\frac{{\bm p}_{\sT}^2}{M_H^2}\, {\bm S}_{\sT}\cdot {\bm s}_{\sT} \right) \right\} \; ,
\label{eq:n.density}
\eea
where $n_- = (1,0,{\bm 0}_{\sT})$. 
It is easy to verify that the unpolarized parton distribution is obtained by 
summing the number densities for all possible four combinations of transverse 
polarizations: $f_1^q = f_{q\uparrow /H\uparrow} + f_{q\downarrow /H\uparrow} 
+ f_{q\uparrow /H\downarrow} + f_{q\downarrow /H\downarrow}$. Similarly, the Sivers 
function emerges as the unbalance between the probabilities of finding unpolarized 
partons inside hadrons transversely polarized in opposite directions, namely
\be
(f_{q\uparrow / p\uparrow} + f_{q\downarrow / p\uparrow}) - 
(f_{q\uparrow / p\downarrow} + f_{q\downarrow / p\downarrow}) = 
f_{q/p\uparrow} - f_{q/p\downarrow} = - f_{1\sT}^{\perp\, q}(x,{\bm p}_{\sT})\, 
\frac{\hat{\bm P}\times {\bm p}_{\sT} \cdot {\bm S}_{\sT}}{M_H} \; .
\label{eq:n.densitySivers}
\ee
The Boer-Mulders function and the transversity emerge from the other two combinations
\bea
(f_{q\uparrow / p\uparrow} + f_{q\uparrow / p\downarrow}) - 
(f_{q\downarrow / p\uparrow} + f_{q\downarrow / p\downarrow}) &= &
f_{q\uparrow/p} - f_{q\downarrow/p} = - h_{1\sT}^{\perp\, q}(x,{\bm p}_{\sT})\, 
\frac{\hat{\bm P}\times {\bm p}_{\sT} \cdot {\bm s}_{\sT}}{M_H} \; ,\label{eq:n.densityBM} \\
(f_{q\uparrow / p\uparrow} - f_{q\uparrow / p\downarrow}) - 
(f_{q\downarrow / p\uparrow} - f_{q\downarrow / p\downarrow}) &= &
h_1^q(x,{\bm p}_{\sT})\, {\bm S}_{\sT}\cdot {\bm s}_{\sT} + 
h_{1\sT}^{\perp\, q}(x,{\bm p}_{\sT})\, \left( 
\frac{{\bm p}_{\sT}\cdot {\bm S}_{\sT}\,{\bm p}_{\sT}\cdot {\bm s}_{\sT}}{M_H^2} - 
\frac{{\bm p}_{\sT}^2}{M_H^2}\,{\bm S}_{\sT}\cdot {\bm s}_{\sT} \right) \; .
\label{eq:n.densityall}
\eea


\section{Monte Carlo generation of Drell-Yan events}
\label{sec:mc}

The implementation of Eq.~(\ref{eq:xsect}) in the Monte Carlo generator is  
represented as~\cite{Bianconi:2004wu}:
\be
\frac{d\sigma}{d\Omega dx_1 dx_2 d{\bm q}_{\sT}} = K \, \frac{1}{s}\, 
|{\cal T}({\bm q}_{\sT}, x_1, x_2, M)|^2 \, \sum_{i=1}^3\, c_i ({\bm q}_{\sT}, 
x_1,x_2) \, S_i(\theta, \phi, \phi_{_{S_2}}) \; ,
\label{eq:mc-xsect}
\ee
where the event distribution is driven by the elementary unpolarized annihilation 
$|{\cal T}|^2$. We assume a factorized transverse-momentum dependence in each parton 
distribution such as to break the convolution ${\cal F}$ in Eq.(\ref{eq:convol}), 
leading to
\be
|{\cal T}|^2 \approx A(q_{\sT},x_1,x_2,M) \, F(x_1,x_2) \; ,
\label{eq:factorized}
\ee
where $q_{\sT} \equiv |{\bm q}_{\sT}|$. The function $A$ is parametrized and
normalized as in Ref.~\cite{Conway:1989fs}, where high-energy Drell-Yan $\pi - p$ 
collisions were considered. The average transverse momentum turns out to be 
$\langle q_{\sT} \rangle > 1$ GeV/$c$, which effectively reproduces the influence 
of sizable QCD corrections beyond the parton model picture of Eq.~(\ref{eq:xsect}). 
It is well known~\cite{Altarelli:1979ub} that such corrections induce also large 
$K$ factors and an $M$ scale dependence in parton distributions, determining their
evolution. As in our previous 
works~\cite{Bianconi:2004wu,Bianconi:2005yj,Bianconi:2005bd,Bianconi:2006hc}, we 
conventionally assume in Eq.~(\ref{eq:mc-xsect}) that $K=2.5$, but we stress that 
in an azimuthal asymmetry the corrections to the cross sections in the numerator 
and in the denominator should approximately compensate each other, as it turns out 
to happen at RHIC~\cite{Martin:1998rz} and JPARC~\cite{Kawamura:2007ze} c.m. 
energies. Since the range of $M$ values here explored is close to the one of 
Ref.~\cite{Conway:1989fs}, where the parametrization of $A, F,$ and $c_i$ in 
Eq.~(\ref{eq:mc-xsect}) was deduced assuming $M$-independent parton distributions, 
we keep our same previous 
approach~\cite{Bianconi:2004wu,Bianconi:2005yj,Bianconi:2005bd,Bianconi:2006hc} and 
use
\be
F(x_1,x_2) = \frac{\alpha^2}{12 Q^2}\,\sum_q\,e_q^2\,
f_1^q(x_1; \bar{q}/H_1) \, f_1^q (x_2; q/H_2) + (\bar{q} \leftrightarrow q) \; , 
\label{eq:mcF}
\ee
where the unpolarized distribution $f_1^q (x)$ for various flavors $q=u,d,s,$ is 
taken again from Ref.~\cite{Conway:1989fs}, unless an explicit different choice is 
mentioned.

The whole solid angle $(\theta, \phi)$ of the final lepton pair in the 
Collins-Soper frame is randomly distributed in each variable. The explicit form 
for sorting it in the Monte-Carlo 
is~\cite{Bianconi:2004wu,Bianconi:2005yj,Bianconi:2006hc}
\bea
\sum_{i=1}^3\, c_i (q_{\sT},x_1,x_2) \, S_i(\theta, \phi, \phi_{S_2}) &= 
&(1 + \cos^2 \theta) + \frac{\nu (x_1,x_2,q_{\sT})}{2}\, \sin^2\theta \, \cos 2\phi 
\nn \\
& &+ |{\bm S}_{2\sT}|\, c_3 (q_{\sT},x_1,x_2)\, S_3 (\theta, \phi, \phi_{S_2}) \; .
\label{eq:mcS}
\eea
If quarks were massless, the virtual photon would be only transversely polarized 
and the angular dependence would be described by the coefficient 
$c_1 = 1, \, S_1 = 1+\cos^2 \theta$. In the following, we discuss violations of such 
azimuthally symmetric dependence, as they appear in Eq.~(\ref{eq:mcS}).


\subsection{The $\cos 2\phi$ asymmetry}
\label{sec:lamtung}

Azimuthal $\cos 2\phi$ asymmetries in Eq.~(\ref{eq:mc-xsect}), induced by 
$c_2 \equiv \textstyle{\frac{\nu}{2}}$, are due to the longitudinal polarization of 
the virtual photon and to the fact that quarks have an intrinsic transverse momentum 
distribution, leading to the explicit violation of the socalled Lam-Tung sum 
rule~\cite{Conway:1989fs}. QCD corrections influence $\nu$, which in principle depends 
also on $M^2$~\cite{Conway:1989fs}. The coefficient $c_2$ was simulated at the GSI 
kinematics~\cite{Bianconi:2004wu}, using the simple parametrization of 
Ref.~\cite{Boer:1999mm} and testing it against the previous measurement of 
Ref.~\cite{Guanziroli:1987rp}. The ${\bm p}_{\sT}$ dependence of the parton 
distributions in the nucleon was parametrized as~\cite{Boer:1999mm} 
\bea
f_1^q(x,{\bm p}_{\sT}) &= &\frac{\alpha_{\sT}}{\pi}\,e^{-\alpha_{\sT}\,{\bm p}_{\sT}^2}\,
f_1^q (x) \nn \\
h_1^{\perp\,q}(x,{\bm p}_{\sT}) &= &c^q\, 
\frac{M_{_C}\, M_N}{{\bm p}_{\sT}^2 +M_{_C}^2} \, f_1^q (x,{\bm p}_{\sT}) \; ,
\label{eq:boerparams}
\eea
where $\alpha_{\sT} = 1$ GeV$^{-2}$, $M_{_C}=2.3$ GeV, $c^q = 1$, and $M_N$ is the 
nucleon mass. Following the steps in Sec.~VI of Ref.~\cite{Boer:1999mm}, one easily 
gets
\bea
c_2^{\nu} &\equiv &\frac{\nu}{2} = 
\frac{\sum_q \, e_q^2 \, {\cal F}\left[ \left( 2 \hat{\bm h}\cdot {\bm p}_{1\sT} \, 
\hat{\bm h}\cdot {\bm p}_{2\sT}-{\bm p}_{1\sT}\cdot {\bm p}_{2\sT} \right) \, 
\displaystyle{\frac{\bar{h}_1^{\perp\,q}(x_1,{\bm p}_{1\sT}) \, 
h_1^{\perp\,q}(x_2,{\bm p}_{2\sT}}{M_1\,M_2}}\, \right] + (1 \leftrightarrow 2)}
{\sum_q\,e_q^2\,{\cal F}\left[ \bar{f}_1^q(x_1,{\bm p}_{1\sT})\, 
f_1^q(x_2,{\bm p}_{2\sT}) \right] + (1 \leftrightarrow 2)} \nonumber \\
&\approx &\frac{4M_C^2 \, q_{\sT}^2}{(q_{\sT}^2+4M_C^2)^2} \; .
\label{eq:nu}
\eea
Since the ${\bm p}_{\sT}$ dependence of $h_1^{\perp}$ was fitted to the measured 
$\cos 2\phi$ asymmetry of the corresponding unpolarized Drell-Yan cross section, 
which is small for $1\lesssim q_{\sT} \lesssim 3$ GeV/$c$ (see, for example, Fig.4 in 
Ref.~\cite{Boer:1999mm}), correspondingly, the $\cos 2\phi$ asymmetry turned out 
to be small for the considered statistically relevant $q_{\sT}$ range in kinematics 
conditions reachable at GSI~\cite{Bianconi:2004wu}.  

In the literature, there are several models available for $h_1^\perp$ in the nucleon; 
in particular, in the context of the spectator diquark model both 
T-even~\cite{Jakob:1997wg} and T-odd parton densities~\cite{Brodsky:2002rv} have been 
analyzed, and further work is in progress~\cite{Gamberg:2007??}. Here, we reconsider 
the spectator diquark model of 
Ref.~\cite{Bacchetta:2003rz} and we present some preliminary results, obtained in the 
context of an improved and enlarged approach~\cite{futuro}. The main features 
of the latter are the following. The gauge link appearing in the quark-quark 
correlator is expanded and truncated at the one-gluon exchange level 
(see Fig.~\ref{fig:link}). The emerging interference diagrams produce the naive 
T-odd structures necessary to have nonvanishing TMD distributions like the 
Sivers $f_{1\sT}^\perp$ and the Boer-Mulders $h_1^\perp$ functions. Because of the 
symmetry properties of the nucleon wave function, scalar and axial-vector diquarks 
are considered with different couplings to the valence quark left. A dipole-like form 
factor is attached at each nucleon-quark-diquark vertex (but other choices will be 
explored~\cite{futuro}). The obtained expressions, that replace the ones in 
Eq.~(\ref{eq:boerparams}), are
\bea
f_1^q(x,{\bm p}_{\sT}) &= &\frac{N_{qS}^2}{16\pi^3}\,
\frac{\left[ {\bm p}_{\sT}^2 + (m+M_Nx)^2 \right]\,(1-x)^3}{\left(A_S^2+{\bm p}_{\sT}^2 
\right)^4} + \frac{N_{qa}^2}{16\pi^3}\,
\frac{\left[ {\bm p}_{\sT}^2 (1+x^2) + (m+M_Nx)^2 (1-x)^2 \right]\,(1-x)}
{\left(A_a^2+{\bm p}_{\sT}^2 \right)^4} \nonumber \\
h_1^{\perp\,q}(x,{\bm p}_{\sT}) &= &\frac{N_{qS}^2}{64\pi^4}\,e e_D\,
\frac{M_N(m+M_Nx)\,(1-x)^3}{A_S^2\, \left(A_S^2+{\bm p}_{\sT}^2 \right)^3} + 
\frac{N_{qa}^2}{64\pi^4}\,e e_D \, 
\frac{M_N (m+M_Nx)\, (1-x)^2}{A_a^2\,\left( A_a^2+{\bm p}_{\sT}^2 \right)^3} \; ,
\label{eq:diquark}
\eea
where 
\be
A_{S/a}^2 = (1-x) \Lambda^2 + x M_{S/a}^2 - x (1-x) M_N^2 \; , \quad 
e e_D = -4\pi\,C_F\,\alpha_s \approx - \frac{16 \pi}{3}\, 0.3 \; ,
\label{eq:cutoff}
\ee
and $m=0.36$ GeV is the valence (constituent) quark mass, $\Lambda =0.5$ is a cutoff 
for high quark virtualities, $M_S=0.6$ GeV and $M_a=0.8$ GeV are the scalar and 
axial-vector diquark masses, respectively. With these choices, the normalizations 
turn out to be $N_{uS}^2=6.52 \times 3/2, \, N_{ua}^2 = 30.63 \times 1/2,\, 
N_{dS}^2=0, \, N_{da}^2= 30.63$. The sign of the obtained $h_1^{\perp\, q}$ (and also 
of $f_{1\sT}^{\perp \, q}$, see below) is consistent with the lattice 
predictions~\cite{Gockeler:2006zu} for both $q=u,d$.

\begin{figure}[h]
\centering
\includegraphics[width=7cm]{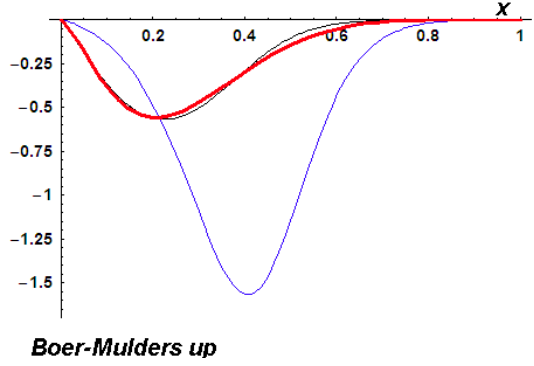}
\caption{The spectator diquark model calculation of the Boer-Mulders function 
$x h_1^{\perp\, u}(x)$ at two different scales: initial $Q_0^2=0.1$ GeV$^2$ (blue 
line), final $Q^2=16$ GeV$^2$ (black line). DGLAP evolution is considered at NLO 
level. The red line is a 3-parameter fit to the black line (see text).}
\label{fig:h1perp}
\end{figure}

As a preliminary step in the analysis of azimuthal asymmetries at COMPASS, for the 
evolution of the parton densities in this context we made an educated guess, since 
this information is unknown for both $f_{1\sT}^{\perp}$ and $h_1^{\perp}$. We have 
integrated the ${\bm p}_{\sT}$ dependence away in Eq.~(\ref{eq:diquark}), and applied 
DGLAP evolution at NLO using the code of Ref.~\cite{Miyama:1995bd} for 
unpolarized quarks and of Ref.~\cite{Hirai:1997mm} for transversely polarized ones. 
The resulting expressions have been inserted in Eq.~(\ref{eq:boerparams}) in place 
of the corresponding $f_1^q (x)$, in order to have an evolved model 
dependence in the behaviour in $x$, but retaining the phenomenological behaviour 
in ${\bm p}_{\sT}$ fitted to the available data. Consistently with the discussed 
approximations, also the evolution scales have been determined in an approximate 
way. The final scale obviously reflects the foreseen COMPASS setup, where 
Drell-Yan dileptons can be detected with invariant masses above 4 GeV; hence, DGLAP 
evolution has been stopped at $Q^2=16$ GeV$^2$. The initial soft scale has been 
deduced by solving the renormalization group equations for the second Mellin moment 
of the non-singlet (valence) distribution, with the anomalous dimension consistently 
determined at NLO; the result is $Q_0^2 \approx 0.1$ GeV$^2$. In 
Fig.~\ref{fig:h1perp} the model $x h_1^{\perp\, u}(x)$ is shown at $Q_0^2$ (blue line) 
and at $Q^2$ (black line); the red line is a fit to the latter one with the form 
$N x^\alpha (1-x)^\beta$, which is actually used in the Monte Carlo calculation of 
the coefficient $c_2^{\nu}$ to produce the azimuthal asymmetry.

\subsection{The Sivers effect}
\label{sec:sivers}

If we consider the Sivers effect in Eq.~(\ref{eq:xsect}), the last term in 
Eq.~(\ref{eq:mcS}) becomes 
\be
S_3^{\mathrm{Siv}} (\theta, \phi, \phi_{S_2}) = (1+\cos^2 \theta) \, 
\sin (\phi - \phi_{S_2}) 
\label{eq:mcS3-sivers}
\ee
and the corresponding coefficient $c_3$ reads
\be
c_3^{\mathrm{Siv}} (q_{\sT},x_1,x_2) = 
\frac{\sum_q\,e_q^2\,{\cal F}\left[ \hat{\bm h}\cdot {\bm p}_{2\sT} \, 
\displaystyle{\frac{f_1^q(x_1,{\bm p}_{1\sT})\, 
f_{1\sT}^{\perp\, q}(x_2,{\bm p}_{2\sT})}{M_2}} \right]}
{\sum_q\,e_q^2\,{\cal F}\left[ f_1^q(x_1,{\bm p}_{1\sT})\, 
 f_1^q(x_2,{\bm p}_{2\sT}) \right]} \; ,
\label{eq:mcc3-sivers}
\ee
where the complete dependence of the involved TMD parton distributions has been
made explicit. For the Sivers function $f_{1\sT}^\perp$, we adopt the 
parametrization described in Ref.~\cite{Bianconi:2005yj}. It is inspired to the 
one of Ref.~\cite{Vogelsang:2005cs}, where the transverse momentum of the detected 
pion in the SIDIS process was assumed to come entirely from the ${\bm p}_{\sT}$ 
dependence of the Sivers function, and was further integrated out building the 
fit in terms of specific moments of the function itself. The $x$ dependence of 
that approach is here retained, but a different flavor-dependent normalization 
and an explicit ${\bm p}_{\sT}$ dependence are introduced that are bound to the 
shape of the recent RHIC data on $pp^\uparrow \to \pi X$ at $\sqrt{s}=200$ 
GeV~\cite{Adler:2005in}, where large persisting asymmetries are found that 
could be partly due to the leading-twist Sivers mechanism. The expression adopted 
is
\bea
f_{1\sT}^{\perp\, q}(x,{\bm p}_{\sT}) &=&N_q\,x\,(1-x)\,
\frac{M_2 p_0^2 p_{\sT}}{(p_{\sT}^2+\frac{p_0^2}{4})^2}\,f_1^q(x,{\bm p}_{\sT}) \nn \\
&= & N_q\,x\,(1-x)\,\frac{M_2 p_0^2 p_{\sT}}{(p_{\sT}^2+\frac{p_0^2}{4})^2}\,
\frac{1}{\pi \, \langle p_{\sT}^2 \rangle}\, e^{-p_{\sT}^2/\langle p_{\sT}^2 \rangle}\, 
f_1^q(x) \; ,
\label{eq:Siversnoi}
\eea
where $p_0 = 2$ GeV/$c$, and $N_u = - N_d = 0.7$. The sign, positive for $u$
quarks and negative for the $d$ ones, already takes into account the predicted sign
change of $f_{1\sT}^\perp$ from Drell-Yan to SIDIS~\cite{Collins:2002kn}. 

Following the steps described in Sec.~III-2 of Ref.~\cite{Bianconi:2005yj}, we 
can directly insert Eq.~(\ref{eq:Siversnoi}) into Eq.~(\ref{eq:mcc3-sivers}) and get
\be
c_3^{\mathrm{Siv}} \approx x_2 \, (1-x_2)\, 
\left( \frac{2\, p_0 \, q_{\sT}}{q_{\sT}^2+p_0^2} \right)^2 \, \frac{8\, N_u + N_d}{9} 
\; .
\label{eq:c3mc-noi}
\ee
The $q_{\sT}$ shape and peak position are in agreement with a similar analysis of the 
azimuthal asymmetry of the unpolarized Drell-Yan data (the violation of the Lam-Tung 
sum rule~\cite{Boer:1999mm}), but are induced also by the observed 
$x_{_F}-q_{\sT}$ correlation in the above mentioned RHIC data for 
$pp^\uparrow \to \pi X$, when it is assumed that the SSA is entirely due to the
Sivers mechanism. This suggests that the maximum asymmetry is reached in 
the upper valence region such that $x_{_F} \approx x_2 \sim \langle q_{\sT} \rangle /
5$~\cite{Adler:2005in}.


\subsection{The Boer-Mulders effect}
\label{sec:BM}

If we consider the Boer-Mulders effect in Eq.~(\ref{eq:xsect}), the 
last term in Eq.~(\ref{eq:mcS}) becomes 
\be
S_3^{\mathrm{BM}} (\theta, \phi, \phi_{S_2}) = \sin^2 \theta \, 
\sin (\phi + \phi_{S_2}) 
\label{eq:mcS3-boer}
\ee
and the corresponding coefficient $c_3$ reads
\be
c_3^{\mathrm{BM}} (q_{\sT},x_1,x_2) = - 
\frac{\sum_q\,e_q^2\,{\cal F}\left[ \hat{\bm h}\cdot {\bm p}_{1\sT} \, 
\displaystyle{\frac{h_1^{\perp\, q}(x_1,{\bm p}_{1\sT})\, 
h_1^q(x_2,{\bm p}_{2\sT})}{M_1}} \right]}
{\sum_q\,e_q^2\,{\cal F}\left[ f_1^q(x_1,{\bm p}_{1\sT})\, 
 f_1^q(x_2,{\bm p}_{2\sT}) \right]} \; .
\label{eq:mcc3-boer}
\ee

As in Ref.~\cite{Boer:1999mm}, one can use Eq.~(\ref{eq:boerparams}) and the 
following parametrization for transversity, 
\be
h_1^q(x,{\bm p}_{\sT}) = \frac{\alpha_{\sT}}{\pi}\,e^{-\alpha_{\sT}\,{\bm p}_{\sT}^2}\,
h_1^q (x) \; ,
\label{eq:h1}
\ee
to get 
\bea
c_3^{\mathrm{BM}} &= &-\frac{2 M_{_C}\,q_{\sT}}{q_{\sT}^2+4 M_{_C}^2}\, 
\frac{\sum_q\,e_q^2\,f_1^q(x_1; \bar{q}/H_1)\,h_1^q(x_2; q/H_2^\uparrow)+ 
      (\bar{q} \leftrightarrow q)}
     {\sum_q\,e_q^2\,f_1^q(x_1; \bar{q}/H_1)\,f_1^q(x_2; q/H_2)+ 
      (\bar{q} \leftrightarrow q)} \nn \\
&\approx &-\frac{2 M_{_C}\,q_{\sT}}{q_{\sT}^2+4 M_{_C}^2} \, 
\frac{f(x_1; \langle \bar{q} \rangle/H_1) \, h_1(x_2;\langle q \rangle/H_2^\uparrow)}
     {f(x_1; \langle \bar{q} \rangle/H_1) \, f_1(x_2;\langle q \rangle/H_2)}  \equiv 
- \frac{2 M_{_C}\,q_{\sT}}{q_{\sT}^2+4 M_{_C}^2} \, 
\frac{h_1(x_2;\langle q \rangle/H_2^\uparrow)}{f_1(x_2;\langle q \rangle/H_2)} \; ,
\label{eq:mcc3-boer2}
\eea
where the second step is justified by assuming that the contribution of each flavor 
to each parton distribution can be approximated by a corresponding average 
function. This approximation, together with the systematic neglect of sea (anti)quark 
contributions, has been widely explored in Ref.~\cite{Bianconi:2006mf}, and it turns 
out to be largely justified at COMPASS kinematics. Because of the lack of 
parametrizations for $h_1^{\perp}$, numerical simulations of $c_3^{\mathrm{BM}}$ 
were performed in Ref.~\cite{Bianconi:2006hc} by making two different guesses for 
the ratio $h_1(x_2;\langle q \rangle/H_2^\uparrow) / f_1(x_2;\langle q \rangle/H_2)$, 
namely the ascending function $\sqrt{x_2}$ and the descending one $\sqrt{1-x_2}$, 
that both respect the Soffer bound. The goal was to determine the minimum number of 
events (compatible with the kinematical setup and cuts) required to produce azimuthal 
asymmetries that could be clearly distinguished like the corresponding originating 
distributions.

Alternatively, here we will show preliminary results for the $c_3^{\mathrm{BM}}$ 
asymmetry obtained in the spectator diquark model~\cite{futuro} by using 
the same strategy discussed in Sec.~\ref{sec:lamtung}. Namely, we start from the 
${\bm p}_{\sT}$-integrated model expressions for $f_1^q$ and $h_1^{\perp\, q}$ from 
Eq.~(\ref{eq:diquark}), complemented by the model transversity
\be
h_1^q(x) = \frac{N_{qS}^2}{48\pi^2}\,
\frac{(m+M_Nx)^2 \,(1-x)^3}{A_S^6} - \frac{N_{qa}^2}{48\pi^2}\, \frac{x(1-x)}{A_a^4}
\; .
\label{eq:diquarkh1}
\ee
Next, we evolve these distributions from $Q_0^2=0.1$ GeV$^2$ to $Q^2=16$ GeV$^2$ with 
a (polarized) DGLAP NLO kernel. Then, we fit each result with different 3-parameter 
forms $N_i x^{\alpha_i} (1-x)^{\beta_i}, \,i=1-3$, and we replace them in the 
corresponding $x$-dependent part of the parametrizations~(\ref{eq:boerparams}) and 
(\ref{eq:h1}). The calculation of $c_3^{\mathrm{BM}}$ then develops in the same way 
up to the first line of Eq.~(\ref{eq:mcc3-boer2}), where each $x$-dependent parton 
density is now replaced by its properly-evolved corresponding expression in the 
spectator diquark model, with no need to apply the flavor average approximation. 
Note that the $f_1^q$ in the numerator now does not simplify with the one in 
the denominator, because the former and the latter are replaced by the 
${\bm p}_{\sT}$-integrated and evolved $h_1^{\perp\, q}(x,{\bm p}_{\sT})$ and 
$f_1^q(x,{\bm p}_{\sT})$ functions of Eq.~(\ref{eq:diquark}), respectively.


\section{Results of Drell-Yan Monte Carlo simulations}
\label{sec:dyssa}

In the Monte Carlo simulation, events are generated by Eq.~(\ref{eq:factorized}), and 
azimuthal asymmetries are produced by Eq.~(\ref{eq:nu}) for the $\cos 2\phi$ 
asymmetry in unpolarized Drell-Yan, by Eq.~(\ref{eq:c3mc-noi}) for the Sivers effect, 
and by Eq.~(\ref{eq:mcc3-boer2}) for the Boer-Mulders effect, respectively, using 
phenomenological parametrizations or model inputs for the involved (TMD) parton 
densities, as it is described in the previous section. The general strategy
is to divide the event sample in two groups, one for positive values "$U$" of 
$S_2 = \sin^2 \theta \cos 2\phi$ in the unpolarized cross section (or of 
$S_3^{\mathrm{Siv}} = (1+\cos^2\theta ) \sin (\phi-\phi_{S_2})$ for Sivers effect, or 
$S_3^{\mathrm{BM}} = \sin^2\theta \sin (\phi+\phi_{S_2})$ for the Boer-Mulders 
effect), and another one for negative values "$D$", then taking the ratio 
$(U-D)/(U+D)$~\cite{Bianconi:2004wu}. Data are accumulated only in the $x_2$ bins of 
the polarized proton, i.e. they are summed over in the $x_1$ bins for the beam, in 
the transverse momentum $q_{\sT}$ of the muon pair and in their zenithal orientation 
$\theta$. Statistical errors for the asymmetry $(U-D)/(U+D)$ are obtained by 
making 10 independent repetitions of the simulation for each individual case, and 
then calculating for each $x_2$ bin the average asymmetry value and the variance. We 
checked that 10 repetitions are a reasonable threshold to have stable numbers, since 
the results do not change significantly when increasing the number of repetitions 
beyond 6.

The transversely polarized proton target is obtained from a $NH_3$ molecule where 
each $H$ nucleus is fully transversely polarized and the number of "polarized" 
collisions is 25\% of the total number of collisions, i.e. with a dilution factor 
1/4~\cite{Bianconi:2004wu}. The beam is represented by charged pions or 
antiprotons such that $100\leq s\leq 200$ GeV$^2$, i.e. roughly the c.m. energy available 
at GSI in the socalled asymmetric collider mode~\cite{Bianconi:2004wu}. The reason 
for employing antiproton beams is simply due to the lack of knowledge of 
$h_1^\perp$ in the pion. Antiproton-proton collisions involve annihilations of 
valence parton densities, as do the pion-proton ones. Hence, it is anyway useful to 
study the shape and size of this asymmetry in the same kinematic conditions, even 
if this kind of projectile is presently not considered at COMPASS. The invariant mass 
of the lepton pair (usually, muons) is constrained in the $1.5<M<2.5$ GeV 
and $4<M<9$ GeV ranges, according to the value of $s$, in order to explore 
approximately the same $x$ valence range and to avoid overlaps with the resonance 
regions of the $\bar{c}c$ and $\bar{b}b$ quarkonium systems. 

Proper cuts are applied to the $q_{\sT}$ distribution, namely $1<q_{\sT} <3$ GeV/$c$, in 
order to filter out low-$q_{\sT}$ muon pairs from background processes other than the 
Drell-Yan mechanism, and to optimize the ratio between the absolute sizes of the 
asymmetry and the statistical errors. The resulting $\langle q_{\sT} \rangle 
\sim 1.8$ GeV/$c$ is in fair agreement with the one experimentally explored at 
RHIC~\cite{Adler:2005in}. Whenever the $\theta$ angular dependence is represented 
by a $\sin^2 \theta$ function, like for the $S_2$ of Eq.~(\ref{eq:mcS}) and for 
the $S_3^{\mathrm{BM}}$ of Eq.~(\ref{eq:mcS3-boer}), the angular distribution is 
constrained in the range $60^{\rm o}< \theta < 120^{\rm o}$, because outside these 
limits the azimuthal asymmetry is too small~\cite{Bianconi:2004wu}. 


\begin{table}
\caption{\label{tab:lumi} Total absorption cross sections per nucleon and related 
Drell-Yan event counts per month, for pion- and antiproton-proton collisions at 
$s=200$ GeV$^2$ and luminosity ${\cal L}=4 \times 10^{31}$ cm$^{-2}$s$^{-1}$, 
producing Drell-Yan pairs for various invariant masses (see text for a discussion 
of the kinematics and the cutoffs).}
\begin{ruledtabular}
\begin{tabular}{cccc}
beam & $M$ (GeV) & $\sigma_{tot}$ (nb/nucleon) & events per month \\
\hline
$\pi$ & 2.5-4 (no $J/\psi$) & 0.5 & $52\,000$ \\
$\pi$ & 4-9 & 0.25 & $26\,000$ \\
$\bar{p}$ & 1.5-2.5 & 2.4 & $249\,600$ \\
$\bar{p}$ & 4-9 & 0.1 & $10\,400$ \\
\end{tabular}
\end{ruledtabular}
\end{table}

The above cuts typically produce a reduction factor of ~2.5 in the initial event 
sample~\cite{Bianconi:2004wu}. For pion beams, we can easily think of samples as 
large as $250\,000$ events, which are then reduced to $100\,000$, and further to 
$25\,000$ in polarized collisions by the target dilution factor. It is much 
more difficult to get these numbers using antiproton beams. However, in our 
simulation of $\bar{p}p^{\left( \uparrow \right)} \to \mu^+ \mu^- X$ processes we 
will keep them in order to make consistent statistical comparisons with the pion 
beam case. In Tab.\ref{tab:lumi}, we list several values of the total absorption 
cross section $\sigma_{\mathrm{tot}}$ per single nucleon, as they are deduced from 
our Monte Carlo for pion and antiproton beams hitting a $NH_3$ target at $s=200$ 
GeV$^2$ and in the above specified conditions, and producing final Drell-Yan muons 
with different invariant masses. At the luminosity ${\cal L}=4 \times 10^{31}$ 
cm$^{-2}$s$^{-1}$ and with efficiency 1, the product $\sigma_{tot} {\cal L}$ gives 
the number of Drell-Yan events per nucleon and per second that could be ideally 
reached at COMPASS. For example, with pion beams of 100 GeV energy hitting a 
transversely polarized $NH_3$ target at $s=200$ GeV$^2$, and producing muon pairs 
with invariant masses in the range 4-9 GeV, it is possible to collect around 
$25\,000$ Drell-Yan events within approximately one month of run at the luminosity 
$4 \times 10^{31}$ cm$^{-2}$s$^{-1}$. 

\begin{figure}[h]
\centering
\includegraphics[width=9cm]{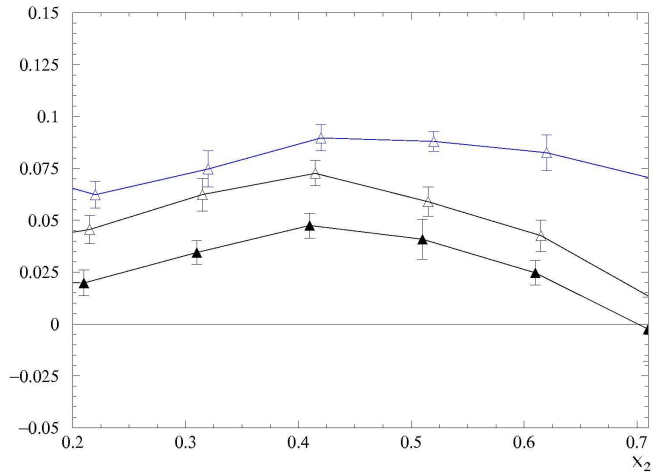}
\caption{The asymmetry $(U-D)/(U+D)$ with positive ($U$) and negative ($D$) values of
$\cos 2\phi$ in Eq.~(\protect{\ref{eq:mcS}}) (see text), for Drell-Yan events from 
the $\bar{p} p \to \mu^+ \mu^- X$ reaction at $\sqrt{s}\sim 14$ GeV, $4<M<9$ GeV, and 
$60^{\rm o}< \theta < 120^{\rm o}$. Blue open triangles for the parametrization of 
Eq.~(\protect{\ref{eq:boerparams}}), constrained to reproduce the experimental data 
of Ref.~\protect{\cite{Guanziroli:1987rp}} and with $2<q_{\sT} <3$ GeV/$c$. Black open 
and closed triangles for the results of the spectator diquark 
model~(\protect{\ref{eq:diquark}}) with $2<q_{\sT} <3$ GeV/$c$ and $1<q_{\sT} <3$ 
GeV/$c$, respectively. Lines are drawn to guide the eyes.}
\label{fig:nu}
\end{figure}

\subsection{The $\cos 2\phi$ asymmetry}
\label{sec:mclamtung}

We first consider the sample of $100\,000$ Drell-Yan events for the 
$\bar{p} p \to \mu^+ \mu^- X$ reaction at $\sqrt{s} \sim 14$ GeV, surviving the cuts 
in the muon invariant mass, $4<M<9$ GeV, and in the angular window,  
$60^{\rm o}< \theta < 120^{\rm o}$. Moreover, there is no target dilution factor;  
this sample can be collected in approximately 10 months of dedicated run (see 
Tab.~\ref{tab:lumi}). Events are accumulated in the $x_2$ bins of the target proton 
and for each bin two groups of events are stored, one corresponding to positive 
values ($U$) of $S_2 = \sin^2 \theta \cos 2\phi$ in Eq.~(\ref{eq:mcS}), and one for 
negative values ($D$). 

In Fig.~\ref{fig:nu}, the asymmetry $(U-D)/(U+D)$ is shown for each bin 
$x_2$. Average asymmetries and (statistical) error bars are obtained by 10 
independent repetitions of the simulation. Boundary values of $x_2$ beyond 0.7 are 
excluded because of very low statistics. Blue open triangles indicate the asymmetry 
generated by $c_2^\nu$ of Eq.~(\ref{eq:nu}), using the parametrization of 
Eq.~(\ref{eq:boerparams}), constrained to reproduce the experimental data 
of Ref.~\cite{Guanziroli:1987rp} and with the further cut $2<q_{\sT} <3$ GeV/$c$. 
Black triangles show the same asymmetry when starting from the parton 
densities of Eq.~(\ref{eq:diquark}) in the spectator diquark model, as explained 
in Sec.~\ref{sec:lamtung}. In this case, open triangles refer to simulations with 
the same cut $2<q_{\sT} <3$ GeV/$c$ as before, while close triangles use the wider 
cut $1<q_{\sT} <3$ GeV/$c$.

The negligible statistical error bars are due to the large sample, which is probably 
unrealistic for a $\bar{p}$ beam, and the displayed sensitivity to kinematical cuts 
in $q_{\sT}$ would also not be reachable in reality. But the bulk message is a 
confirmation of the findings in Refs.~\cite{Boer:1999mm} and \cite{Bianconi:2004wu}: 
since the ${\bm p}_{\sT}$ dependence of $h_1^{\perp}$ is fitted (also for the 
spectator diquark model) to the measured $\cos 2\phi$ asymmetry of the unpolarized 
Drell-Yan cross section in the experimental conditions of 
Ref.~\cite{Guanziroli:1987rp}, which is small for $1\lesssim q_{\sT} \lesssim 3$ 
GeV/$c$, the resulting asymmetry in the COMPASS conditions turns out unavoidably 
small, as it was the case for the simulation at the GSI 
kinematics~\cite{Bianconi:2004wu}. Still, the asymmetry itself is nonvanishing and 
deserves to be measured in order to attack the problem of building a parametrization 
for $h_1^\perp$.

\subsection{The Sivers effect}
\label{sec:mcsivers}

In order to study the Sivers effect, we consider two samples, one of $100\,000$ 
Drell-Yan events for the $\pi^- p^\uparrow \to \mu^+ \mu^- X$ reaction again at 
$\sqrt{s} \sim 14$ GeV and for muon invariant mass in the $4<M<9$ GeV range, and 
another one of $25\,000$ events for the $\pi^+ p^\uparrow \to \mu^+ \mu^- X$ reaction 
in the same kinematic conditions. Both samples can be accumulated approximately in 
the same time. As before, the transverse momentum distribution is constrained in the 
range $1<q_{\sT} <3$ GeV/$c$, the samples are collected in $x_2$ bins and for each bin 
two groups of events are stored, one corresponding to positive values ($U$) of 
$\sin (\phi - \phi_{S_2})$ in Eq.~(\ref{eq:mcS3-sivers}), and one for negative values 
($D$). 

\begin{figure}[h]
\centering
\includegraphics[width=9cm]{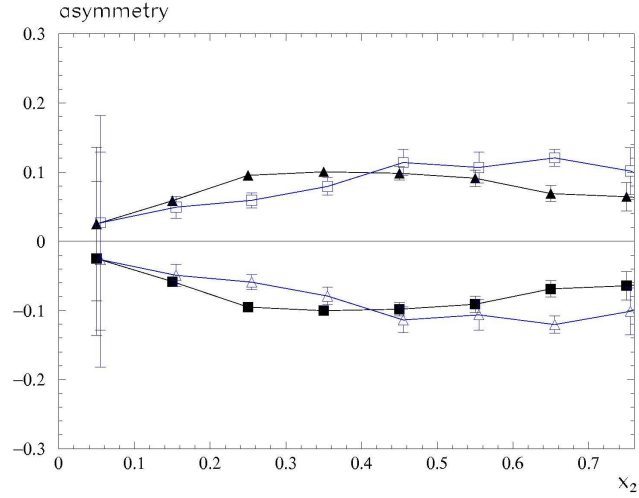}
\caption{The asymmetry $(U-D)/(U+D)$ with positive ($U$) and negative ($D$) values of
$\sin (\phi - \phi_{S_2})$ in Eq.~(\protect{\ref{eq:mcS3-sivers}}) (see text), for 
Drell-Yan events from the Sivers effect in the $\pi^\pm p^\uparrow \to \mu^+ \mu^- X$ 
reaction at $\sqrt{s}\sim 14$ GeV, $4<M<9$ GeV, and $1<q_{\sT} <3$ GeV/$c$, using 
the parametrization of Eq.~(\protect{\ref{eq:Siversnoi}}) for the Sivers function 
(see text). Triangles indicate $100\,000$ events collected in $x_2$ bins for the 
$\pi^-$ beam and with $N_u>0$; squares for $N_u<0$. Open triangles indicate 
$25\,000$ events for the $\pi^+$ beam and with $N_u>0$; open squares for $N_u<0$. 
Lines are drawn to guide the eye.}
\label{fig:sivers}
\end{figure}

In Fig.~\ref{fig:sivers}, the asymmetry $(U-D)/(U+D)$ is shown for each bin 
$x_2$. Average asymmetries and (statistical) error bars are obtained, as usual, by 10 
independent repetitions of the simulation. Boundary values of $x_2$ beyond 0.7 are 
excluded because of very low statistics. The triangles indicate the results with 
the $\pi^-$ beam obtained by Eq.~(\ref{eq:c3mc-noi}) assuming that 
$f_{1\sT}^\perp$ changes sign from SIDIS to Drell-Yan~\cite{Collins:2002kn}. For sake 
of comparison, the squares illustrate the opposite results that one would obtain by 
ignoring such prediction. Finally, the open triangles and open squares refer to the 
same situation, respectively, but for the $\pi^+$ beam. 

In the valence picture of the $(\pi^-)\pi^+-p$ collision where the $(\bar{u}u) \, 
\bar{d}d$ annihilation dominates, the SSA for the Drell-Yan process induced by 
$\pi^+$ has opposite sign with respect to $\pi^-$ because of the opposite signs for 
the normalization $N_f, \, f=u,d$, in the parametrization~(\ref{eq:Siversnoi}). 
Apart for very low $x_2$ values where the parton picture leading to 
Eq.~(\ref{eq:xsect}) becomes questionable, the error bars are very small and allow 
for a clean reconstruction of the asymmetry shape and, more importantly, for a 
conclusive test of the predicted sign change in $f_{1\sT}^\perp$.

\begin{figure}[h]
\centering
\includegraphics[width=6.5cm]{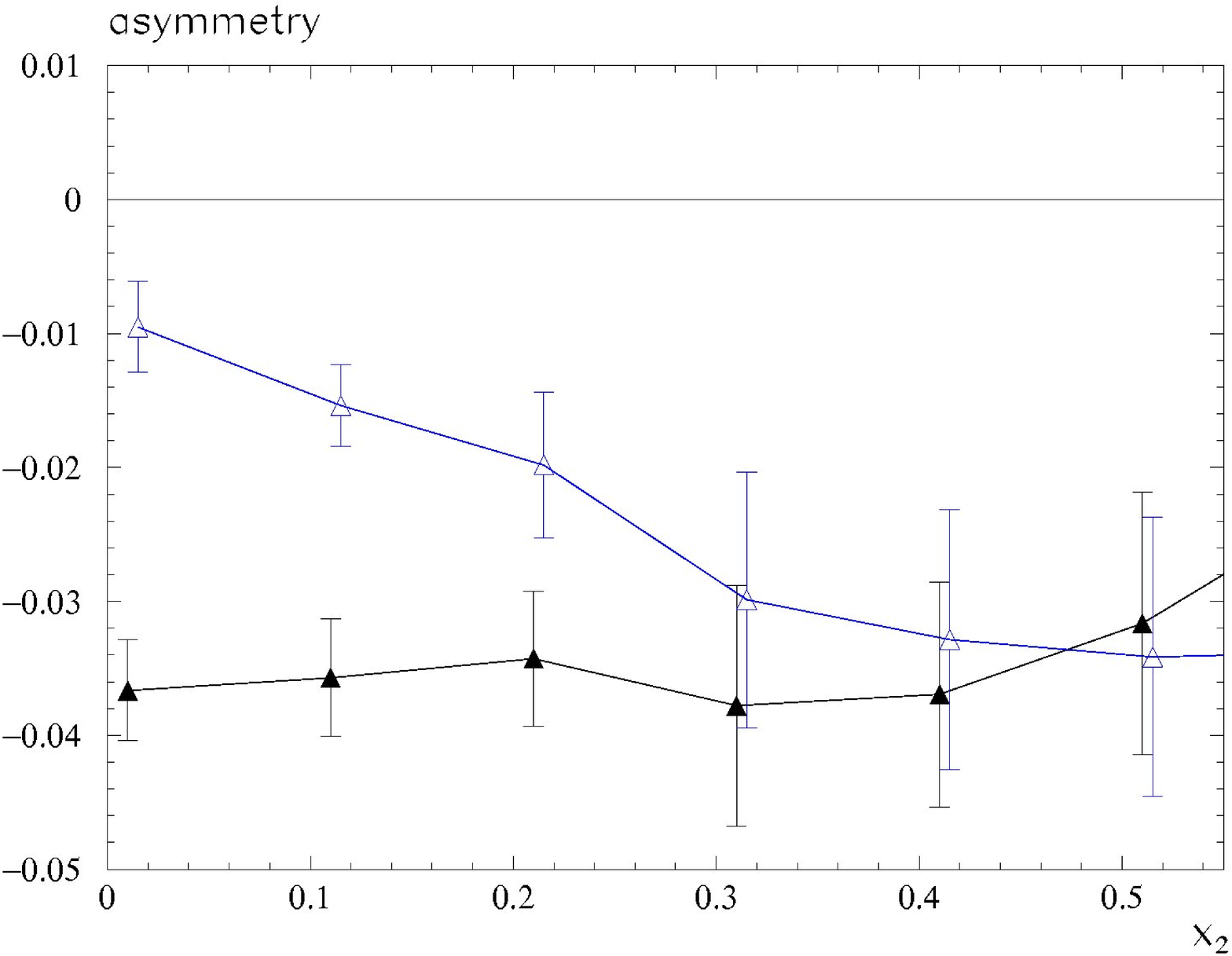}\hspace{.5cm}
\includegraphics[width=6.5cm]{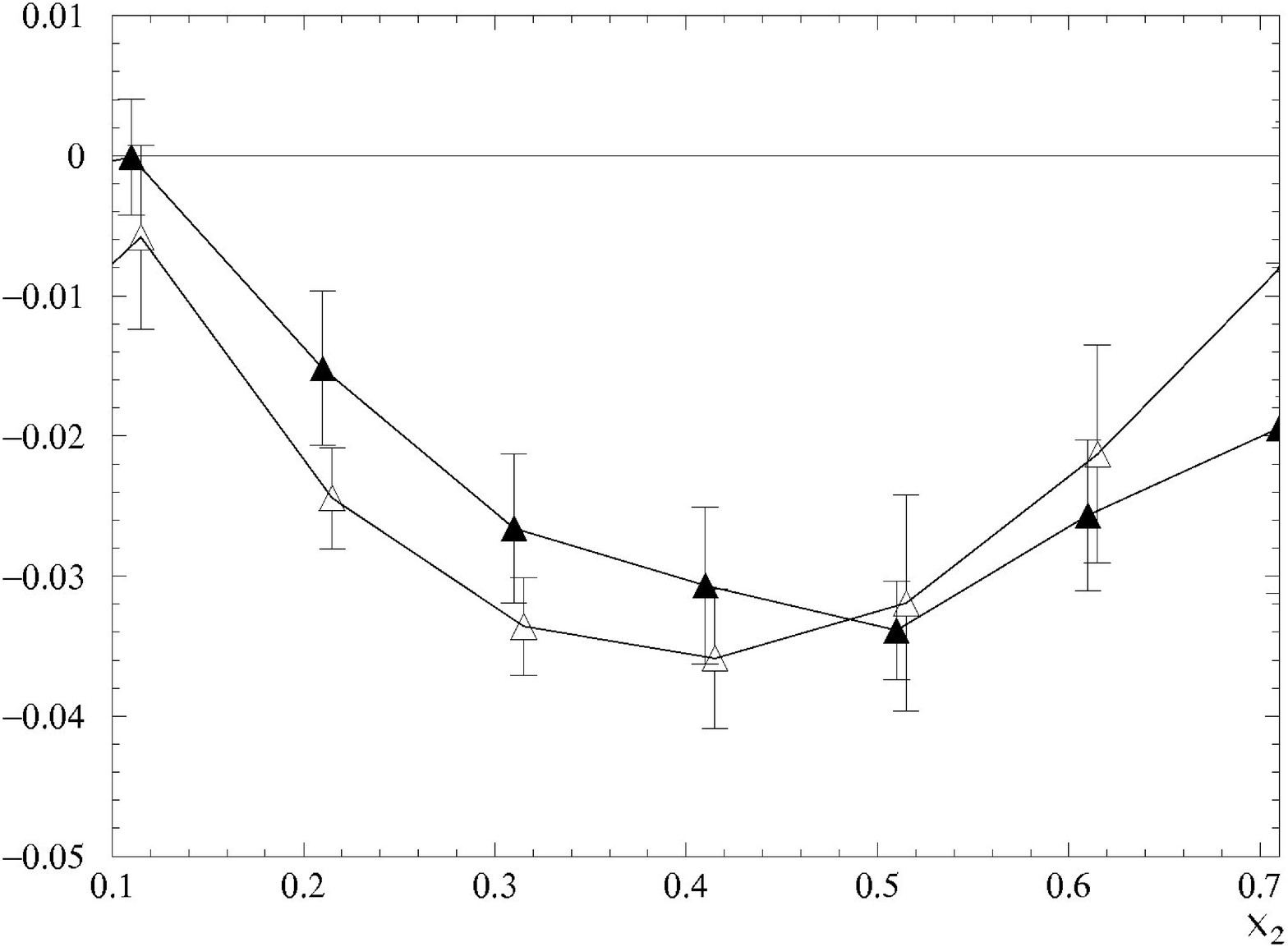}
\caption{The asymmetry $(U-D)/(U+D)$ with positive ($U$) and negative ($D$) values of
$\sin (\phi + \phi_{S_2})$ in Eq.~(\protect{\ref{eq:mcS3-boer}}) (see text), for 
$200\,000$ Drell-Yan events from the Boer-Mulders effect. In the left panel, the 
$\pi^- p^\uparrow \to \mu^+ \mu^- X$ reaction is considered at $\sqrt{s}\sim 10$ 
GeV, $1.5<M<2.5$ GeV, and $1<q_{\sT} <3$ GeV/$c$; triangles for 
$h_1(x_2, \langle q \rangle / H_2^\uparrow) / f_1 (x_2, \langle q \rangle / H_2) = 
\sqrt{1-x_2}$ inside Eq.~(\protect{\ref{eq:mcc3-boer2}}) (see text), open triangles 
for $h_1(x_2, \langle q \rangle / H_2^\uparrow) / f_1 (x_2, \langle q \rangle / H_2) 
= \sqrt{x_2}$; the $\langle q \rangle$ notation represents a common average term that 
replaces each contribution in the flavor sum (for further details, see text). In 
the right panel, the $\bar{p} p^\uparrow \to \mu^+ \mu^- X$ reaction is considered 
at $\sqrt{s}\sim 14$ GeV, $4<M<9$ GeV; triangles for the results from the spectator 
diquark model with $1<q_{\sT} <3$ GeV/$c$, open triangles for $2<q_{\sT} <3$ GeV/$c$. 
In all cases, the further cut $60^{\rm o}< \theta < 120^{\rm o}$ is applied. Lines 
are drawn to guide the eye.}
\label{fig:BM}
\end{figure}

\subsection{The Boer-Mulders effect}
\label{sec:mcBM}

For the Boer-Mulders effect, we have collected $200\,000$ Drell-Yan events for the 
$\pi^- p^\uparrow \to \mu^+ \mu^- X$ reaction at the lower $\sqrt{s} \sim 10$ GeV 
in order to keep a significant statistics; the muon invariant mass is restricted in 
the $1.5<M<2.5$ GeV range, with $1<q_{\sT} <3$ GeV/$c$ and 
$60^{\rm o}< \theta < 120^{\rm o}$. In the left panel of Fig.~\ref{fig:BM}, the 
asymmetry $(U-D)/(U+D)$ is shown for each bin $x_2$ between the events with positive 
$(U)$ and negative $(D)$ values of $\sin (\phi + \phi_{S_2})$ in 
Eq.~(\ref{eq:mcS3-boer}). Triangles correspond to the choice 
$h_1(x_2, \langle q \rangle / H_2^\uparrow) / f_1 (x_2, \langle q \rangle / H_2) = 
\sqrt{1-x_2}$ inside Eq.~(\ref{eq:mcc3-boer2}), open triangles to the choice 
$h_1(x_2, \langle q \rangle / H_2^\uparrow) / f_1 (x_2, \langle q \rangle / H_2) = 
\sqrt{x_2}$ (see Sec.~\ref{sec:BM}). Both choices respect the Soffer bound between 
$h_1$ and $f_1$. The error bars are only statistical and are small because of 
the copious statistics. As it is evident in the figure, for $x_2 < 0.3$ it is 
possible to distinguish the trend of the triangles (which statistically reflects the 
descending trend of the input function $\sqrt{1-x_2}$) from the one of the open 
triangles (referred to the ascending $\sqrt{x_2}$). We made simulations also at the 
same kinematics of the previous Sivers effect, namely for $\sqrt{s} \sim 14$ GeV and 
for muon pair invariant masses in the $4<M<9$ GeV range. The statistical error bars 
remain small and the SSA is definitely nonvanishing, but a clear distinction between 
the two trends is no longer possible. 

In the right panel of Fig.~\ref{fig:BM}, the same $(U-D)/(U+D)$ asymmetry is 
considered for the $\bar{p} p^\uparrow \to \mu^+ \mu^- X$ reaction at 
$\sqrt{s} \sim 14$ GeV with muon invariant mass in the $4<M<9$ GeV range and 
$60^{\rm o}< \theta < 120^{\rm o}$. The same sample of $200\,000$ events is now 
produced by input parton densities from the spectator diquark model. Triangles 
correspond to the events with the cut $1<q_{\sT} <3$ GeV/$c$, open triangles with 
$2<q_{\sT} <3$ GeV/$c$. Again, boundary values of $x_2$ beyond 0.7 are excluded because 
of very low statistics. Within the (statistical) error bars, there is no sensitivity 
to the cut in the lepton pair transverse momentum. 

The overall size of the spin asymmetry is small and the reached statistical 
accuracy indicates that the size of the sample is not responsible for this feature 
(see also Ref.~\cite{Sissakian:2005yp}). Rather, it is the outcome of a combination 
of several sources. First of all, the phenomenological ${\bm p}_{\sT}$ dependence of 
$h_1^{\perp}$ in Eq.~(\ref{eq:boerparams}), which also affects the result with the 
spectator diquark model according to the approximation described in 
Sec.~\ref{sec:BM}: it is fixed to reproduce the data of 
Ref.~\cite{Guanziroli:1987rp}, which indicate significant asymmetries only for very 
large ${\bm p}_{\sT}$ beyond the range of interest here. Secondly, the target dilution 
factor: it is unavoidably introduced by the features of the actual target that 
effectively reproduces a transversely polarized proton; at COMPASS the wanted 
"polarized" events are 1/4 of the total number of collisions on the $NH_3$ molecule. 
Finally, the Soffer bound: the various choices for the ratio 
$h_1(x_2, \langle q \rangle / H_2^\uparrow) / f_1 (x_2, \langle q \rangle / H_2)$ on 
one side, as well as the expressions~(\ref{eq:diquark}) and (\ref{eq:diquarkh1}) for  
$f_1, h_1^\perp,$ and $h_1$ in the spectator diquark model on the other side, are
tightly limited by this constraint.


\section{Two-hadron inclusive production in hadronic collisions}
\label{sec:ppDiFF}

As anticipated in Sec.~\ref{sec:intro}, the most popular technique to extract the 
transversity $h_1$ is to measure a single-spin asymmetry (SSA) in SIDIS production 
of pions on transversely polarized proton targets. The parton fragmentation into the 
detected pion can be described by a quark-quark correlator similar to the one 
of Eq.~(\ref{eq:phiT}), namely
\bea
\Delta (z,{\bm k}_{\sT}) &= &\frac{1}{2z}\,\int dk^+ \, \sum_{\scriptstyle X}\,\int 
\frac{d^4\xi}{(2\pi )^4}\,e^{ik\cdot \xi}\,
\langle 0|U_{\left[ \infty, \xi \right]}\, \psi (\xi)|P_h, X\rangle 
\,\langle P_h, X|\bar{\psi}(0) \, U_{\left[ 0,\infty \right]} 
|0\rangle\Big\vert_{k^-=P_h^- / z} \nonumber \\
&= &\frac{1}{2P_h^-}\, \left\{ D_1(z,{\bm k}_{\sT}) \, \Pslash_h + H_1^\perp(z,
{\bm k}_{\sT}) \, \sigma_{\mu \nu} \, \frac{k^\mu P_h^\nu}{M_h} \right\} \; , 
\label{eq:deltaT}
\eea
which holds, in general, for a transversely polarized parton with momentum $k$,  
fragmenting in an unpolarized hadron with momentum $P_h$ and mass $M_h$. Applying the 
same definition of projection as in Eq.~(\ref{eq:phiGamma}), and following again the 
prescriptions of Ref.~\cite{Bacchetta:2004jz},  we can define the number density 
$D_{h/q s_{T}}$ of unpolarized hadrons $h$ in a parton with flavor $q$ and 
transverse polarization ${\bm s}_{\sT}$:
\bea
D_{h / q s_{T}} &= &\frac{1}{2}\, 
\left( \Delta^{[ \gamma_\mu \, n_{\scriptscriptstyle +}^\mu / 2 ]} + 
\Delta^{[ i\sigma_{\mu \nu}\, n_{\scriptscriptstyle +}^\mu s_{\scriptscriptstyle T}^\nu 
\gamma_5 /2 ]}  \right) \nonumber \\
&= &\frac{1}{2}\,\left\{ D_1^q(z,{\bm k}_{\sT}) + H_1^{\perp\, q}(z,{\bm k}_{\sT})\, 
\frac{\hat{\bm k}\times {\bm P}_{h\sT}\cdot {\bm s_{\sT}}}{z M_h} \right\} 
\; , 
\label{eq:n.density2}
\eea
where $n_+=(0,1,{\bm 0}_{\sT})$. It is evident that the unpolarized "decay" function 
is given by the sum upon all possible polarizations of the fragmenting quark, namely 
$D_1^q = D_{h/q\uparrow}+D_{h/q\downarrow}$, while the Collins function is deduced by  
\be
D_{h/q\uparrow} - D_{h/q\downarrow} = H_1^{\perp\, q}(z,{\bm k}_{\sT})\, 
\frac{\hat{\bm k}\times {\bm P}_{h\sT}\cdot {\bm s_{\sT}}}{z M_h} \; .
\label{eq:n.densityCollins}
\ee

Combining the quark-quark correlators of Eq.~(\ref{eq:phiT}) and (\ref{eq:deltaT}), 
together with the elementary cross section for the virtual photon absorption, it is 
possible to build the leading-twist expression of the SIDIS cross section for 
a transversely polarized target in a factorized picture (for the general expression, 
see Ref.~\cite{Bacchetta:2006tn}). The transversity $h_1$ can be extracted by the 
following SSA:
\be
\frac{\displaystyle{\int d\phi_{\scriptstyle S} d{\bm P}_{h\sT} 
\frac{|{\bm P}_{h\sT}|}{M_h} \, \sin (\phi_h+\phi_{\scriptstyle S})\, 
(d\sigma^\uparrow - d\sigma^\downarrow)}} 
{\displaystyle{\int d\phi_{\scriptstyle S} d{\bm P}_{h\sT} \, (d\sigma^\uparrow + 
d\sigma^\downarrow)}} \propto 
\frac{\sum_{q\bar{q}} \, e^2_q \, h_1^q(x) \, H_1^{\perp q \, (1)} (z)}
{\sum_{q\bar{q}} \, e^2_q \, f_1^q(x) \, D_1^q(z)} \; , 
\label{eq:collins_ssa}
\ee
where $\phi_{\scriptstyle S}, \phi_h,$ are the azimuthal orientations with respect to 
the scattering plane of the target polarization vector and of the hadronic plane 
containing ${\bm P}_{h\sT}$, respectively. From Eq.~(\ref{eq:collins_ssa}) it is 
evident that this strategy, known as Collins effect~\cite{Collins:1993kk}, requires 
the knowledge of the entire ${\bm P}_{h\sT}$ distribution of the produced hadron $h$, 
which is obviously not possible from the experimental point of view. But also 
theoretically there are some complications, since the factorization proof and the 
evolution equations of TMD parton density functions are more 
involved~\cite{Collins:2004nx,Ji:2004wu}. For the case of hadronic collisions
leading to semi-inclusive final states in specific kinematic conditions, very recently 
an explicit factorization-breaking mechanism has been discussed for a ${\bm
k}_{\sT}$-dependent elementary hard cross section~\cite{Collins:2007nk}. 

\begin{figure}[h]
\centering
\includegraphics[width=5cm]{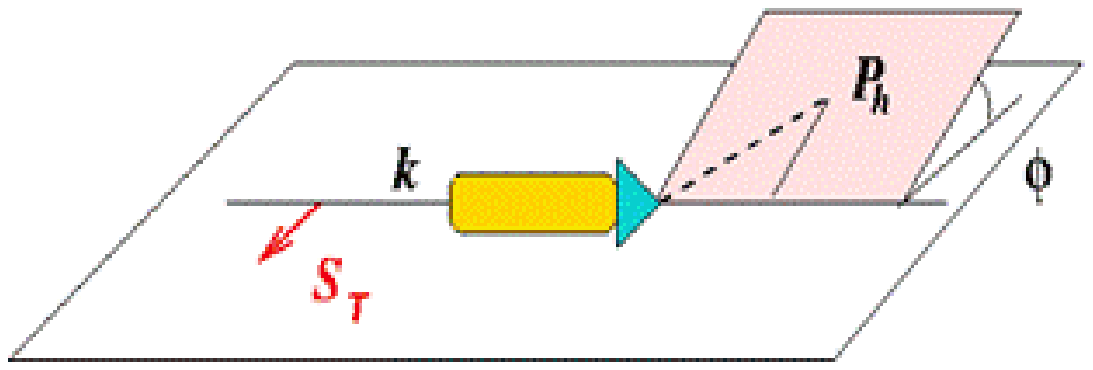}\hspace{.5cm}
\includegraphics[width=5cm]{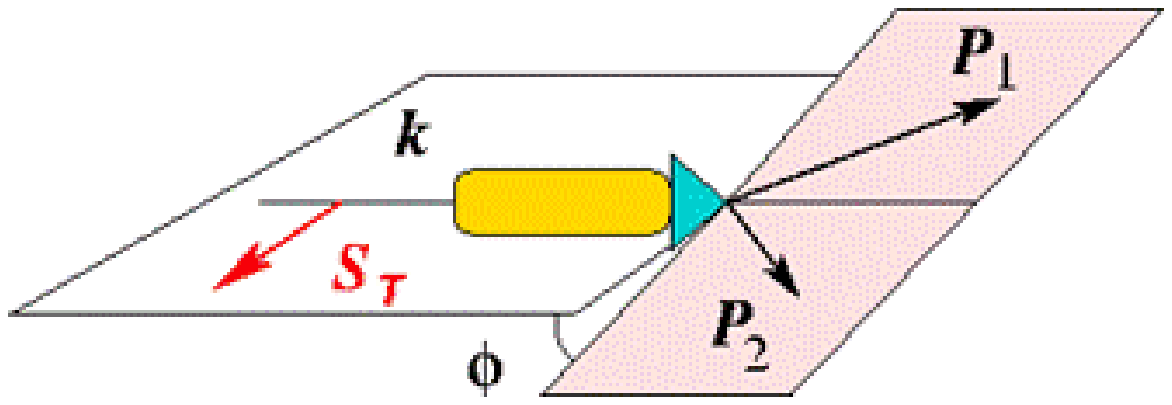}
\caption{In the left panel, the graphic representation of the mixed product 
${\bm k} \times {\bm P}_{h\sT} \cdot {\bm s}_{\sT}$, which governs the Collins effect. 
In the right panel, the mixed product 
${\bm P}_1 \times {\bm P}_2 \cdot {\bm s}_{\sT}$, which describes the asymmetric 
distribution in $\phi$ of a hadron pair collinear to the jet axis $\hat{\bm k}$.}
\label{fig:mixedproducts}
\end{figure}

In Fig.~\ref{fig:mixedproducts}, the Collins effect produced by the mixed product 
${\bm k} \times {\bm P}_{h\sT} \cdot {\bm s}_{\sT}$ (left panel), is compared with the 
more exclusive situation where two hadrons are 
produced~\cite{Collins:1994kq,Jaffe:1998hf}. In this case (right panel), it is 
possible to relate the polarization ${\bm s}_{\sT}$ of the fragmenting quark to the 
azimuthal orientation of the plane containing the two hadrons via the mixed product 
${\bm P}_1 \times {\bm P}_2 \cdot {\bm s}_{\sT}$. If $P_h=P_1+P_2$ is the total 
momentum of the pair, and $2R = P_1 - P_2$ the relative one, the asymmetry survives 
even if the pair is taken collinear to the fragmenting quark direction $\hat{\bm k}$, 
or alternatively if the dependence upon ${\bm P}_{h\sT}$ is integrated away, contrary 
to what happens in the Collins effect. In fact, the quark-quark correlator for the 
fragmentation of a transversely polarized quark into two unpolarized hadrons reads
\bea
\Delta(z,\zeta,{\bm k}_{\sT},{\bm R}_{\sT}^2,\phi_{\scriptstyle R}) &=&
\frac{1}{8z\,(1-\zeta^2)}\, \int dk^+\, \sum_{\scriptstyle X} \, \int 
\frac{d^4\xi}{(2\pi)^{4}}\, e^{+i k \cdot \xi} \langle 0| 
{\cal U}_{(-\infty,\xi)}\,\psi(\xi)|P_h, R; X\rangle \nonumber \\
& &\mbox{\hspace{6truecm}} \times \, \langle P_h, R; X| 
\bar{\psi}(0)\, {\cal U}_{(0,-\infty)} | 0 \rangle\,\Big\vert_{k^-=P_h^-/z} 
\nonumber \\
&= &\frac{1}{16\pi} \left\{ D_1(z,\zeta,{\bm k}_{\sT},{\bm R}_{\sT}^2) + i 
\frac{\Rslash_{\sT}}{M_h}\,  H_1^{\open}(z,\zeta,{\bm k}_{\sT},{\bm R}_{\sT}^2) 
\right. \nonumber \\
& &\left. + i \frac{\kslash_{\sT}}{M_h} \, H_1^\perp(z,\zeta,{\bm k}_{\sT},{\bm R}_{\sT}^2)
+ \frac{\epsilon_{\sT}^{ij} k_{\sT i} R_{\sT j}}{M_h^2} \gamma_5 \, 
G_1^\perp(z,\zeta,{\bm k}_{\sT},{\bm R}_{\sT}^2) \right\} \, \gamma^+ \; , 
\label{eq:delta2T}
\eea
where $z=P_h^-/k^-=z_1+z_2$ is the fraction of quark energy delivered to the pair, 
$\zeta = 2R^-/P_h^-=(z_1-z_2)/z$ describes how this fraction is split inside the 
pair, and $\epsilon_{\sT}^{ij} \equiv \epsilon^{-+ij}$. The new fragmentation functions 
appearing in Eq.~(\ref{eq:delta2T}) are named Dihadron Fragmentation Functions 
(DiFF)~\cite{Bianconi:1999cd}. The $H_1^\perp$ is the analogue of the Collins 
function, while $G_1^\perp$ describes the same situation but for the fragmenting 
quark longitudinally polarized (a sort of analogue of the jet handedness, see 
Ref.~\cite{Boer:2003ya} and references therein). After integrating upon 
${\bm k}_{\sT} = - {\bm P}_{h\sT}/z$, both disappear and we are left with
\be
\Delta(z,\zeta,{\bm R}_{\sT}^2,\phi_{\scriptstyle R}) =\frac{1}{16\pi} \left\{ 
D_1(z,\zeta,{\bm R}_{\sT}^2) + i \frac{\Rslash_{\sT}}{M_h}\,  
H_1^{\open}(z,\zeta,{\bm R}_{\sT}^2) \right\} \, \gamma^+ \; .
\label{eq:delta2}
\ee
Apart for the usual "decay" probability $D_1$, a potentially asymmetric term survives 
involving the chiral-odd DiFF $H_1^{\open}$, which appears as the natural partner of 
the transversity $h_1$ in the SSA~\cite{Radici:2001na}
\be
\frac{1}{\sin (\phi_{\scriptstyle R} + \phi_{\scriptstyle S})}\, 
\frac{d\sigma^\uparrow - d\sigma^\downarrow}{d\sigma^\uparrow + d\sigma^\downarrow} 
\propto \frac{|{\bm R}|}{M_h} \, 
\frac{\sum_{q\bar{q}} \, e_q^2\,h_1^q(x)\, H_{1}^{\open\, q}(z,\zeta,{\bm R}_{\sT}^2)}
{\sum_{q\bar{q}} \, e_q^2\,f_1^q(x)\,D_{1}^q(z,\zeta,{\bm R}_{\sT}^2)} \; .
\label{eq:DiFF_ssa}
\ee
The $H_1^\open$ equally describes the fragmentation of a transversely polarized 
quark into two unpolarized hadrons as $H_1^\perp$, but it survives the 
${\bm k}_{\sT}$ integration since it is related to the asymmetry depicted in 
Fig.~\ref{fig:mixedproducts}, which is connected to the ${\bm R}_{\sT}$ vector. In this
respect, any observable related to the asymmetry quantitatively described by
$H_1^{\open}$ is free of the problems mentioned above about the Collins mechanism, in
particular about the ${\bm k}_{\sT}$ factorization. 

It is possible also to expand these DiFF in the relative partial waves of the two 
hadrons, by suitably rewriting the $\zeta$ dependence in their c.m. 
frame~\cite{Bacchetta:2002ux}. Very recently, the HERMES collaboration reported 
measurements of the SSA in Eq.~(\ref{eq:DiFF_ssa}) involving only the interference 
between $s-$ and $p-$wave components, which is necessary to generate na\"ive T-odd 
functions like $H_1^\open$~\cite{vanderNat:2005yf}. The COMPASS collaboration also 
presented analogous preliminary results~\cite{Joosten:2005}. In the meanwhile, 
the BELLE collaboration is planning to measure $H_1^{\open}$ in the near 
future~\cite{Abe:2005zx}. A spectator model calculation of leading-twist DiFF has 
been built by tuning the parameters on the output of pion pair distributions 
(proportional to $D_1$) from PYTHIA, after adapting it to the HERMES 
kinematics~\cite{Bacchetta:2006un}. Evolution equations for DiFF have been also 
studied, including the full dependence upon ${\bm R}_{\sT}^2$, or, equivalently, upon 
the pair invariant mass~\cite{Ceccopieri:2007ip}. 

Here, we reconsider the proposal formulated for the first time in 
Ref.~\cite{Bacchetta:2004it}, namely to use hadronic collisions on a transversely 
polarized proton target and inclusively detect one (or two) pairs of pions. 

\begin{figure}
\centering
\includegraphics[width=9cm]{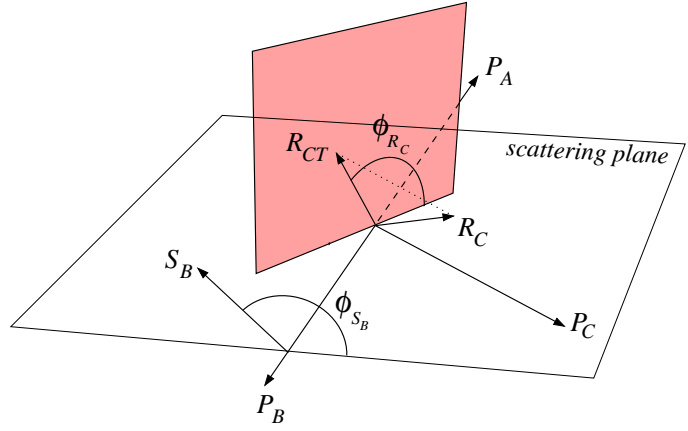}
\caption{Description of the kinematics for the proton-proton annihilation into a 
single pair in a jet.}
\label{fig:ppplanes}
\end{figure}

We consider first the process $A+B\rightarrow (h_1 h_2)_C+X$, where two protons (with 
momenta $P_A, P_B$, and spin vectors $S_A, S_B$) collide, and two unpolarized hadrons 
$h_1, h_2$, with momenta $P_1, P_2$, masses $M_1, M_2,$ and invariant mass squared 
$M_C^2=P_C^2=(P_1+P_2)^2$, are inclusively detected inside the same jet. The 
intrinsic transverse component ${\bm P}_{C\sT}$ with respect to the jet axis is 
integrated over and, consequently, ${\bm P}_C$ is taken parallel to the jet axis 
itself. The component ${\bm P}_{C\perp}$ perpendicular to the beam direction 
(defined by ${\bm P}_A$, see Fig.~\ref{fig:ppplanes}) serves as the hard scale of the 
process; it is assumed to be much bigger than the masses of the colliding hadrons and 
of $M_h$. The following analysis is valid only at leading order in 
$1/|{\bm P}_{C\perp}|$, i.e.\ at leading twist. 

The azimuthal angles are defined in the hadronic center of mass as follows 
(see also Fig.~\ref{fig:ppplanes})
\begin{align} 
\cos \phi_{S_A} &= 
  \frac{(\hat{\bm P}_A \times {\bm P}_C)}{|\hat{\bm P}_A\times{\bm P}_C|}
  \cdot \frac{(\hat{\bm P}_A\times{\bm S}_A)}{|\hat{\bm P}_A
     \times{\bm S}_A|},
&
\sin \phi_{S_A} &= 
  \frac{({\bm P}_C \times {\bm S}_A) \cdot \hat{\bm P}_A}{|\hat{\bm P}_A
     \times{\bm P}_C|\,|\hat{\bm P}_A\times{\bm S}_A|} , \\
\label{eq:azang2}
\cos \phi_{S_B} &= 
  \frac{(\hat{\bm P}_B \times {\bm P}_C)}{|\hat{\bm P}_B\times{\bm P}_C|}
  \cdot \frac{(\hat{\bm P}_B\times{\bm S}_B)}{|\hat{\bm P}_B
     \times{\bm S}_B|},
&
\sin \phi_{S_B} &= 
  \frac{({\bm P}_C \times {\bm S}_B) \cdot \hat{\bm P}_B}{|\hat{\bm P}_B
     \times{\bm P}_C|\,|\hat{\bm P}_B\times{\bm S}_B|} , \\
\label{eq:azang3}
\cos \phi_{R} &= 
  \frac{(\hat{\bm P}_C \times {\bm P}_A)}{|\hat{\bm P}_C\times{\bm P}_A|}
  \cdot \frac{(\hat{\bm P}_C\times{\bm R}_C)}{|\hat{\bm P}_C
     \times{\bm R}_C|},
&
\sin \phi_{R} &= 
  \frac{({\bm P}_A \times {\bm R}_C) \cdot \hat{\bm P}_C}{|\hat{\bm P}_C
     \times{\bm P}_A|\,|\hat{\bm P}_C\times{\bm R}_C|} \; .
\end{align}  

The partons involved in the elementary scattering have momenta 
$p_a = x_a P_A$, $p_b = x_b P_B$, and $p_c = P_C/z_c$. Also the usual Mandelstam 
variables have a counterpart at the elementary partonic level:
\begin{align}
  \label{eq:mandelstam}
  \hat{s} &= x_a x_b\, s, & \hat{t} &= \frac{x_a}{z_c}\,t, & \hat{u}=  
\frac{x_b}{z_c}\,u \; .
\end{align} 
Conservation of momentum at the partonic level implies that
\be 
\begin{split}
  \hat{s}\, \delta(\hat{s}+\hat{t}+\hat{u}) &= z_c\, \delta\left(z_c + \frac{x_a t
    + x_b u}{x_a x_b s}\right) \,\stackrel{\mbox{\tiny c.m.s.}}{=}\,
z_c\, \delta\left(z_c- \frac{|{\bm P}_{C\perp}|}{\sqrt{s}}\, \frac{x_a e^{-\eta_C}
    + x_b e^{\eta_C}}{x_a x_b}\right) \, \equiv \, z_c \, \delta \left( z_c - \bar{z}_c
    \right) \; ,
\end{split}
\label{eq:stu}
\ee
where $\eta_C$ is the pseudorapidity with respect to ${\bm P}_A$. 

Therefore, the cross section can be written as
\be
\begin{split}
\frac{d\sigma}{d\eta_C\, d|{\bm P}_{C\perp}|\, d\zeta_c \, d {\bm R}_{C\sT} \, 
d\phi_{S_A} \,d \phi_{S_B} } & = 2 \, |{\bm P}_{C\perp}|
\sum_{a,b,c,d}\frac{1}{4}\sum_{({\rm all}\; \chi{\rm's})}
\int \frac{dx_a dx_b}{4 \pi^2 z_c} \, \Phi_a(x_a,S_A)_{\chi_a'\chi^{}_a} \, 
\Phi_b(x_b,S_B)_{\chi_b' \chi^{}_b} \\
& \quad\frac{1}{16 \pi \hat{s}^2} \, \hat{M}_{\chi^{}_c, \chi^{}_d; \chi^{}_a,  
\chi^{}_b} \, \hat{M}^{\ast}_{\chi_a', \chi_b'; \chi_c' \chi_d'} \, 
 \Delta_c (\bar{z}_c,\zeta_c, {\bm R}_{C\sT}^2,\phi_{R_C})_{\chi_c' \chi^{}_c}  \, 
 \delta_{\chi_d' \chi^{}_d} \; ,
\end{split}
\label{eq:1pairxsect}
\ee
where the indices $\chi$'s refer to the chiralities/helicities of the partons, and 
the partonic hard amplitudes $\hat{M}(\hat{s},\hat{t},\hat{u})$ can be taken from 
Refs.~\cite{Gastmans:1990xh,Bacchetta:2004it}. In Eq.~(\ref{eq:1pairxsect}), $\Phi$ 
and $\Delta$ are the same parton-parton correlators of Eqs.~(\ref{eq:phiT}) and 
(\ref{eq:delta2T}), after integrating upon the intrinsic tranverse momenta and using 
the parton helicity basis representation. 

For initial quarks with flavor $a$ we have~\cite{Jaffe:1992ra,Bacchetta:1999kz}
\be
  \Phi_a(x_a,S_A)_{\chi_a' \chi^{}_a} = \left( 
    \begin{array}{cc}
    f_1(x_a) + S_{A\, L} \, g_1(x_a) & |{\bm S}_{A\sT}| \, e^{-i\phi_{S_A}} \, 
    h_1(x_a) \\[5pt]
   |{\bm S}_{A\sT}| \, e^{i\phi_{S_A}} \, h_1(x_a) & f_1(x_a) - S_{A\, L} \, 
    g_1(x_a) \end{array} \right) \; .
\label{eq:phia}
\ee
The correlator is written in the helicity basis where ${\bm P}_A$ defines the 
$\hat{z}$ axis, with $S_{A\, L}$ and $|{\bm S}_{A\sT}|$ indicating the parallel and 
transverse components of the polarization ${\bm S}_A$ with respect to ${\bm P}_A$, 
and the $\hat{x}$ axis is oriented along ${\bm P}_{C\perp}$. Similarly for quark $b$ 
in the hadron $B$. When the initial partons are gluons, in a spin-$\half$ target the 
$\Phi$ is diagonal because of angular momentum 
conservation~\cite{Jaffe:1996ik,Mulders:2000sh}. 

If the final fragmenting parton $c$ is a quark, we have
\be
  \Delta_c (z_c,\zeta_c , {\bm R}_{C\sT}^2,\phi_{R_C})_{\chi_c' \chi_c^{}} = 
  \frac{1}{4 \pi} \, 
   \left( \begin{array}{cc}
       D_1 & i e^{i\phi_{R_C}} \, \frac{|{\bm R}_{C\sT}|}{M_C} \, H_1^{\open} \\[5pt]
       -i e^{-i\phi_{R_C}} \, \frac{|{\bm R}_{C\sT}|}{M_C}\, H_1^{\open} & 
       D_1^q \end{array} \right) \; ,
\label{eq:deltahh'}
\ee
where the helicity basis is now choosen with the $\hat{z}$ axis along ${\bm P}_C$ and 
the $\hat{x}$ axis along the component of ${\bm P}_A$ orthogonal to ${\bm P}_C$ 
(see Fig.~\ref{fig:ppplanes}). For final fragmenting gluons, because of angular 
momentum conservation the $\Delta_c$ can be obtained in analogy to the case of gluon 
distributions in spin-1 targets~\cite{Jaffe:1989xy}:
\be
  \Delta_{c=g} (z_c,\zeta_c ,{\bm R}_{C\sT}^2,\phi_{R_C})_{\chi_c' \chi_c^{}} = 
  \frac{1}{4 \pi} \, 
   \left( \begin{array}{cc}
       \hat{G} & i e^{2i\phi_{R_C}}\, \frac{|{\bm R}_{C\sT}|^2}{M_C^2}\, 
       \delta \hat{G}^\open \\[5pt]
       -i e^{-2i\phi_{R_C}}\, \frac{|{\bm R}_{C\sT}|^2}{M_C^2} \,  
       \delta \hat{G}^\open & \hat{G}  \end{array} \right) \; .
\label{eq:deltagluon}
\ee
The functions $\hat{G}$ and $\delta \hat{G}^\open$ describe the decay into two 
unpolarized hadrons of an unpolarized and a transversely polarized gluon, 
respectively, where by "transverse polarization" we mean linear polarization along 
two independent directions transverse to $\hat{z}$, as in the quark 
case~\cite{Bacchetta:2004it}. Unfortunately, $\delta \hat{G}^\open$ cannot appear in 
connection with the quark transversity $h_1$ because of the mismatch in the units of 
helicity flip between a spin-$\half$ and a spin-1 objects, leading to the 
$\exp(i\phi_{R_C})$ and $\exp(2i\phi_{R_C})$ dependences in Eq.~(\ref{eq:phia}) and 
Eq.~(\ref{eq:deltagluon}), respectively. It can be coupled to the gluon transversity 
$\delta G$, but only in a target with spin greater than 
$\half$~\cite{Jaffe:1989xy}. 

When considering the process $pp^\uparrow \to (h_1 h_2) X$, the most interesting SSA 
is 
\be
A_N (\eta_C, |{\bm P}_{C\perp}|, \zeta_c, {\bm R}_{C\sT}^2, \phi_{R_C}, \phi_{S_B})
= \frac{d\sigma^\uparrow - d\sigma^\downarrow}{d\sigma^\uparrow + d\sigma^\downarrow} 
\equiv \frac{d\sigma_{UT}}{d\sigma_{UU}} \; ,
\label{eq:AN}
\ee
where
\bea
d\sigma_{UU} &=  &2 \, |{\bm P}_{C\perp}| \, \sum_{a,b,c,d}\int 
\frac{d x_a d x_b }{8 \pi^2 z_c} \, f_1^a(x_a) \, f_1^b(x_b) \, 
\frac{d\hat{\sigma}_{ab \to cd}}{d\hat{t}} \, D_1(\bar{z}_c,\zeta_c ,{\bm R}_{C\sT}^2) 
\; , \nonumber \\
d\sigma_{UT} &=&2 \, |{\bm P}_{C\perp}|\, \sum_{a,b,c,d}\, |{\bm S}_{B\sT}| \, 
\sin{(\phi_{S_B}-\phi_{R_C})} \int \frac{d x_a d x_b }{8 \pi^2 z_c} \, f_1^a(x_a) \, 
h_1^b(x_b) \, \frac{d\Delta \hat{\sigma}_{ab^\uparrow \to c^\uparrow d}}{d\hat{t}}
\, \frac{|{\bm R}_{C\sT}|}{M_C}\, H_1^{\open}(\bar{z}_c,\zeta_c , {\bm R}_{C\sT}^2)
\; .
\label{eq:sigmaOO+OT}
\eea
Here, it is understood that when the parton is a gluon, we need to replace $f_1$ and 
$g_1$ with the momentum and helicity gluon distributions, respectively, as well as 
to have $D_1 \to \hat{G}$. The elementary cross sections are~\cite{Bacchetta:2004it}
\bea
\frac{\hat{\sigma}_{ab \to cd}}{d\hat{t}} &\equiv &\frac{1}{16 \pi \hat{s}^2}\,
\frac{1}{4}\sum_{({\rm all}\; \chi{\rm's})} \, \hat{M}_{\chi^{}_c, \chi^{}_d; 
\chi^{}_a,  \chi^{}_b} \, \hat{M}^{\ast}_{\chi_a, \chi_b; \chi_c \chi_d} \; , 
\nonumber \\
\frac{d\Delta \hat{\sigma}_{ab^\uparrow \to c^\uparrow d}}{d\hat{t}} &\equiv 
&\frac{1}{16 \pi \hat{s}^2}\,\frac{1}{4}\sum_{({\rm all}\; \chi{\rm's})} \, 
\hat{M}_{\chi^{}_c, \chi^{}_d; \chi^{}_a, \chi^{}_b} \,  
\hat{M}^{\ast}_{\chi_a, -\chi_b; -\chi_c, \chi_d}  \; .
\label{eq:mmstar}
\eea
In the polarized case, a parton $b$ is transversely polarized in a direction forming 
an azimuthal angle $\phi_{S_b}$ around ${\bm P}_B$ and the transverse polarization of 
parton $c$ forms the same azimuthal angle $\phi_{S_c}=\phi_{S_b}$ around 
${\bm P}_C$ ($\phi_{S_b}$ and $\phi_{S_c}$ are defined analogously to $\phi_{S_B}$ 
and $\phi_{R_C}$, respectively, see Fig.~\ref{fig:ppplanes}).

Using Eq.~(\ref{eq:AN}), the extraction of $h_1$ is possible either by using models 
for DiFF~\cite{Bacchetta:2006un}, or by awaiting for $e^+ e^-$ measurements to 
determine DiFF at BELLE~\cite{GrossePerdekamp:2002eb}, in combination with the recent 
one in SIDIS at HERMES~\cite{vanderNat:2005yf}. Experimental results at different 
energies can be presently related through DGLAP equations at NLL accuracy including 
full dependence upon ${\bm R}_{\sT}^2$~\cite{Ceccopieri:2007ip}. However, DiFF 
can be measured independently in the very same proton-proton collisions analyzed 
so far, by simply detecting another hadron pair in the other recoiling back-to-back 
jet. In fact, by generalizing the formalism of Eqs.~(\ref{eq:1pairxsect}) and 
(\ref{eq:sigmaOO+OT}), we have~\cite{Bacchetta:2004it}
\be
d\sigma_{UU}  = {\cal{A}} + \cos{(\phi_{R_C}-\phi_{R_D})} \; {\cal{B}} +  
\cos{(2\phi_{R_C}-2\phi_{R_D})} \; {\cal{C}} \; ,
\label{eq:2pairxsect}
\ee
where
\bea
{\cal{A}} & = &\sum_{a,b,c,d}\int \frac{d x_a}{8 \pi^2} \, f_1^a(x_a) \, 
\bar{x}_b\,f_1^b(\bar{x}_b) \, \frac{d\hat{\sigma}_{ab \to cd}}{d \hat{t}} \, 
D_1(\bar{z}_c,\zeta_c , {\bm R}_{C\sT}^2) \, D_1(\bar{z}_d,\zeta_d ,
{\bm R}_{D\sT}^2) \; , \nonumber \\
{\cal{B}}  &= &\sum_{a,b,c,d}\int  \frac{d x_a}{8 \pi^2} \, f_1^a(x_a) \, 
\bar{x}_b\,f_1^b(\bar{x}_b) \, 
\frac{d\Delta \hat{\sigma}_{ab \to c^\uparrow d^\uparrow}}{d \hat{t}} \,  
\frac{|{\bm R}_{C\sT}|}{M_{C}}\, H_1^{\open}(\bar{z}_c,\zeta_c ,
{\bm R}_{C\sT}^2) \, \frac{|{\bm R}_{D\sT}|}{M_{D}} \, 
H_1^{\open}(\bar{z}_d,\zeta_d ,{\bm R}_{D\sT}^2) \; , \nonumber \\
{\cal{C}}  &=  &\sum_{a,b,c,d}\int  \frac{d x_a}{8 \pi^2}\, f_1^a(x_a) \, 
\bar{x}_b\,f_1^b(\bar{x}_b) \, 
\frac{d\Delta \hat{\sigma}_{ab \to c^\uparrow d^\uparrow}}{d\hat{t}} \, 
\frac{|{\bm R}_{C\sT}|^2}{M_{C}^2}\, \delta \hat{G}^{\open}
(\bar{z}_c,\zeta_c ,{\bm R}_{C\sT}^2)\, \frac{|{\bm R}_{D\sT}|^2}{M_{D}^2}\, 
\delta \hat{G}^{\open}(\bar{z}_d,\zeta_d ,{\bm R}_{D\sT}^2) \; , 
\label{eq:abc}
\eea
and 
\be
\frac{d\Delta \hat{\sigma}_{ab \to c^\uparrow d^\uparrow}}{d\hat{t}} \equiv  
\frac{1}{16 \pi \hat{s}^2}\,\frac{1}{4}\sum_{({\rm all}\; \chi{\rm's})} \, 
\hat{M}_{\chi^{}_c, \chi^{}_d; \chi^{}_a, \chi^{}_b} \, 
\hat{M}^{\ast}_{\chi_a, \chi_b; -\chi_c, -\chi_d} \; ,
\ee
with $\bar{x}_b = x_a\, e^{-\eta_C}\,e^{-\eta_D}$ by momentum conservation. 

All functions ${\cal A}, {\cal B}$ and ${\cal C}$ are interesting. The first two 
contain pairs of the DiFF $D_1, H_1^{\open}$, one for each hadron pair: measuring 
the symmetric and the $\cos{(\phi_{R_C}-\phi_{R_D})}$ asymmetric parts of the cross 
section for the $pp \to (h_1 h_2)_{jet_C} (h_1 h_2)_{jet_D} X$ process, it allows the 
extraction of $D_1$ and $H_1^{\open}$ and, in turn, of the transversity $h_1$ from 
the asymmetry $A_N$ of Eqs.~(\ref{eq:AN},\ref{eq:sigmaOO+OT}) in the corresponding 
polarized collision $pp^\uparrow \to (h_1 h_2) X$. Finally, the ${\cal C}$ 
observable describes the fragmentation of two transversely (linearly) polarized 
gluons into two transversely (linearly) polarized spin-1 resonances, which would not 
be available in a SIDIS process using spin-$\half$ proton targets.


\section{Conclusions}
\label{sec:end}

Using energetic hadronic (pion) beams on a transversely polarized $NH_3$ target, it 
opens a new window at COMPASS for the exploration of the partonic (spin) structure 
of the nucleon. When releasing the condition for collinear elementary annihilations, 
the leading-twist cross section for the production of Drell-Yan muon pairs contains 
several azimuthal asymmetric terms, whose combined study would shed light on yet 
unresolved puzzles. For example, the $\cos 2\phi$ term in the unpolarized cross 
section involves twice the exotic partonic density $h_1^\perp$, the socalled 
Boer-Mulders function, that describes the unbalance in the distribution of 
tranversely polarized partons inside unpolarized spin-$\half$ hadrons. It could give 
a natural explanation of the role of this contribution in the yet unexplained 
violation of the socalled Lam-Tung sum rule. The same $h_1^\perp$ 
happens in the $\sin(\phi + \phi_{_S})$ asymmetric part of the polarized 
cross section, with $\phi_{_S}$ the target polarization orientation. It is 
convoluted with the tranversity $h_1$, then offering an alternative strategy for its 
extraction with respect to the standard Collins effect in SIDIS. Finally, the 
polarized cross section contains also a $\sin(\phi - \phi_{_S})$ asymmetric part 
where the usual momentum density $f_1$ is convoluted with $f_{1\sT}^\perp$, the 
Sivers function, that describes the distortion of unpolarized parton distributions by 
the transverse polarization of the parent hadron. Studying all these partonic 
densities allows to build a 3-dim. map of partons inside hadrons via the 
determination of their angular orbital momentum, contributing to the solution of 
the yet pending puzzle about the spin sum rule of the proton. Moreover, a theorem 
based on very general hypotheses, would predict a sign change for $f_{1\sT}^\perp$ and 
$h_1^\perp$ being extracted from Drell-Yan measurements with respect to a SIDIS one. 
An experimental verification of this represents a formidable test of QCD in its 
nonperturbative domain.

Monte Carlo simulations of the above azimuthal (spin) asymmetries at COMPASS, reveal 
that a very good statistics can be reached in short running time, given the relative 
abundance of energetic pions in the beam and the foreseen high luminosity. As a 
consequence, COMPASS is an ideal place for studying the $\sin(\phi - \phi_{_S})$ 
single-spin asymmetry involving the Sivers function, and for testing its predicted 
sign change with respect to SIDIS. Very small statistical error bars are reachable 
also for the other two asymmetries, but experimental constraints coming from older 
experiments (NA10, E615,..) reduce the size of the simulated asymmetry; a lack of 
parametrizations for the Boer-Mulders function, either for proton targets or for 
pion beams, further makes its extraction an issue under debate. 

We have also shown that in semi-inclusive (transversely polarized) collisions with 
production of one hadron pair in the same jet, it is possible to isolate the 
convolution $f_1\otimes h_1 \otimes H_1^{\open}$, involving the $H_1^\open$ 
describing the fragmentation of a transversely polarized parton in the two observed 
hadrons, through the measurement of the asymmetry of the cross section 
in the azimuthal orientation of the pair around its c.m. momentum.
In the production of two hadron pairs in two separate jets 
in unpolarized collisions, it is possible to isolate the convolution 
$f_1 \otimes f_1 \otimes H_1^{\open}\otimes H_1^{\open}$, through the measurement of 
the asymmetry of the cross section in the relative azimuthal orientation of the two 
pairs. Since no distribution and fragmentation functions with an explicit 
transverse-momentum dependence are required, there is no need to consider suppressed 
contributions in the elementary cross sections included in the convolutions and the 
discussed asymmetries remain at leading-twist. Therefore, contrary to what happens in 
SIDIS, proton-proton collisions offer a unique possibility to determine 
self-consistently all the unknown parton densities that are necessary to extract the 
transversity $h_1$. We believe that this option should be seriously considered at 
COMPASS.


\bibliographystyle{apsrev}
\bibliography{iff,hadron}

\begin{thebibliography}{70}
\expandafter\ifx\csname natexlab\endcsname\relax\def\natexlab#1{#1}\fi
\expandafter\ifx\csname bibnamefont\endcsname\relax
  \def\bibnamefont#1{#1}\fi
\expandafter\ifx\csname bibfnamefont\endcsname\relax
  \def\bibfnamefont#1{#1}\fi
\expandafter\ifx\csname citenamefont\endcsname\relax
  \def\citenamefont#1{#1}\fi
\expandafter\ifx\csname url\endcsname\relax
  \def\url#1{\texttt{#1}}\fi
\expandafter\ifx\csname urlprefix\endcsname\relax\def\urlprefix{URL }\fi
\providecommand{\bibinfo}[2]{#2}
\providecommand{\eprint}[2][]{\url{#2}}

\bibitem[{\citenamefont{Airapetian et~al.}(2005)}]{Airapetian:2004tw}
\bibinfo{author}{\bibfnamefont{A.}~\bibnamefont{Airapetian}}
  \bibnamefont{et~al.} (\bibinfo{collaboration}{HERMES}),
  \bibinfo{journal}{Phys. Rev. Lett.} \textbf{\bibinfo{volume}{94}},
  \bibinfo{pages}{012002} (\bibinfo{year}{2005}), \eprint{hep-ex/0408013}.

\bibitem[{\citenamefont{Diefenthaler}(2005)}]{Diefenthaler:2005gx}
\bibinfo{author}{\bibfnamefont{M.}~\bibnamefont{Diefenthaler}}
  (\bibinfo{year}{2005}), \bibinfo{note}{proceedings of the 13th International
  Workshop on Deep-Inelastic Scattering (DIS05), 27 Apr - 1 May, 2005, Madison
  - Wisconsin}, \eprint{hep-ex/0507013}.

\bibitem[{\citenamefont{Avakian et~al.}(2005)\citenamefont{Avakian, Bosted,
  Burkert, and Elouadrhiri}}]{Avakian:2005ps}
\bibinfo{author}{\bibfnamefont{H.}~\bibnamefont{Avakian}},
  \bibinfo{author}{\bibfnamefont{P.}~\bibnamefont{Bosted}},
  \bibinfo{author}{\bibfnamefont{V.}~\bibnamefont{Burkert}}, \bibnamefont{and}
  \bibinfo{author}{\bibfnamefont{L.}~\bibnamefont{Elouadrhiri}}
  (\bibinfo{collaboration}{CLAS}) (\bibinfo{year}{2005}),
  \bibinfo{note}{proceedings of 13th International Workshop on Deep-Inelastic
  Scattering (DIS 05), 27 Apr - 1 May, 2005, Madison - Wisconsin},
  \eprint{nucl-ex/0509032}.

\bibitem[{\citenamefont{Alexakhin et~al.}(2005)}]{Alexakhin:2005iw}
\bibinfo{author}{\bibfnamefont{V.~Y.} \bibnamefont{Alexakhin}}
  \bibnamefont{et~al.} (\bibinfo{collaboration}{COMPASS}),
  \bibinfo{journal}{Phys. Rev. Lett.} \textbf{\bibinfo{volume}{94}},
  \bibinfo{pages}{202002} (\bibinfo{year}{2005}), \eprint{hep-ex/0503002}.

\bibitem[{\citenamefont{Radici and van~der Steenhoven}(2004)}]{cerncourier}
\bibinfo{author}{\bibfnamefont{M.}~\bibnamefont{Radici}} \bibnamefont{and}
  \bibinfo{author}{\bibfnamefont{G.}~\bibnamefont{van~der Steenhoven}},
  \bibinfo{journal}{CERN Courier} \textbf{\bibinfo{volume}{44}},
  \bibinfo{pages}{51} (\bibinfo{year}{2004}).

\bibitem[{\citenamefont{Bunce et~al.}(1976)}]{Bunce:1976yb}
\bibinfo{author}{\bibfnamefont{G.}~\bibnamefont{Bunce}} \bibnamefont{et~al.},
  \bibinfo{journal}{Phys. Rev. Lett.} \textbf{\bibinfo{volume}{36}},
  \bibinfo{pages}{1113} (\bibinfo{year}{1976}).

\bibitem[{\citenamefont{Adams et~al.}(1991)}]{Adams:1991cs}
\bibinfo{author}{\bibfnamefont{D.~L.} \bibnamefont{Adams}} \bibnamefont{et~al.}
  (\bibinfo{collaboration}{FNAL-E704}), \bibinfo{journal}{Phys. Lett.}
  \textbf{\bibinfo{volume}{B264}}, \bibinfo{pages}{462} (\bibinfo{year}{1991}).

\bibitem[{\citenamefont{Kane et~al.}(1978)\citenamefont{Kane, Pumplin, and
  Repko}}]{Kane:1978nd}
\bibinfo{author}{\bibfnamefont{G.~L.} \bibnamefont{Kane}},
  \bibinfo{author}{\bibfnamefont{J.}~\bibnamefont{Pumplin}}, \bibnamefont{and}
  \bibinfo{author}{\bibfnamefont{W.}~\bibnamefont{Repko}},
  \bibinfo{journal}{Phys. Rev. Lett.} \textbf{\bibinfo{volume}{41}},
  \bibinfo{pages}{1689} (\bibinfo{year}{1978}).

\bibitem[{\citenamefont{Falciano et~al.}(1986)}]{Falciano:1986wk}
\bibinfo{author}{\bibfnamefont{S.}~\bibnamefont{Falciano}} \bibnamefont{et~al.}
  (\bibinfo{collaboration}{NA10}), \bibinfo{journal}{Z. Phys.}
  \textbf{\bibinfo{volume}{C31}}, \bibinfo{pages}{513} (\bibinfo{year}{1986}).

\bibitem[{\citenamefont{Guanziroli et~al.}(1988)}]{Guanziroli:1987rp}
\bibinfo{author}{\bibfnamefont{M.}~\bibnamefont{Guanziroli}}
  \bibnamefont{et~al.} (\bibinfo{collaboration}{NA10}), \bibinfo{journal}{Z.
  Phys.} \textbf{\bibinfo{volume}{C37}}, \bibinfo{pages}{545}
  (\bibinfo{year}{1988}).

\bibitem[{\citenamefont{Conway et~al.}(1989)}]{Conway:1989fs}
\bibinfo{author}{\bibfnamefont{J.~S.} \bibnamefont{Conway}}
  \bibnamefont{et~al.}, \bibinfo{journal}{Phys. Rev.}
  \textbf{\bibinfo{volume}{D39}}, \bibinfo{pages}{92} (\bibinfo{year}{1989}).

\bibitem[{\citenamefont{Anassontzis et~al.}(1988)}]{Anassontzis:1987hk}
\bibinfo{author}{\bibfnamefont{E.}~\bibnamefont{Anassontzis}}
  \bibnamefont{et~al.}, \bibinfo{journal}{Phys. Rev.}
  \textbf{\bibinfo{volume}{D38}}, \bibinfo{pages}{1377} (\bibinfo{year}{1988}).

\bibitem[{\citenamefont{Lam and Tung}(1980)}]{Lam:1980uc}
\bibinfo{author}{\bibfnamefont{C.~S.} \bibnamefont{Lam}} \bibnamefont{and}
  \bibinfo{author}{\bibfnamefont{W.-K.} \bibnamefont{Tung}},
  \bibinfo{journal}{Phys. Rev.} \textbf{\bibinfo{volume}{D21}},
  \bibinfo{pages}{2712} (\bibinfo{year}{1980}).

\bibitem[{\citenamefont{Brandenburg et~al.}(1994)\citenamefont{Brandenburg,
  Brodsky, Khoze, and Muller}}]{Brandenburg:1994wf}
\bibinfo{author}{\bibfnamefont{A.}~\bibnamefont{Brandenburg}},
  \bibinfo{author}{\bibfnamefont{S.~J.} \bibnamefont{Brodsky}},
  \bibinfo{author}{\bibfnamefont{V.~V.} \bibnamefont{Khoze}}, \bibnamefont{and}
  \bibinfo{author}{\bibfnamefont{D.}~\bibnamefont{Muller}},
  \bibinfo{journal}{Phys. Rev. Lett.} \textbf{\bibinfo{volume}{73}},
  \bibinfo{pages}{939} (\bibinfo{year}{1994}), \eprint{hep-ph/9403361}.

\bibitem[{\citenamefont{Eskola et~al.}(1994)\citenamefont{Eskola, Hoyer,
  Vanttinen, and Vogt}}]{Eskola:1994py}
\bibinfo{author}{\bibfnamefont{K.~J.} \bibnamefont{Eskola}},
  \bibinfo{author}{\bibfnamefont{P.}~\bibnamefont{Hoyer}},
  \bibinfo{author}{\bibfnamefont{M.}~\bibnamefont{Vanttinen}},
  \bibnamefont{and} \bibinfo{author}{\bibfnamefont{R.}~\bibnamefont{Vogt}},
  \bibinfo{journal}{Phys. Lett.} \textbf{\bibinfo{volume}{B333}},
  \bibinfo{pages}{526} (\bibinfo{year}{1994}), \eprint{hep-ph/9404322}.

\bibitem[{\citenamefont{Berger and Brodsky}(1979)}]{Berger:1979du}
\bibinfo{author}{\bibfnamefont{E.~L.} \bibnamefont{Berger}} \bibnamefont{and}
  \bibinfo{author}{\bibfnamefont{S.~J.} \bibnamefont{Brodsky}},
  \bibinfo{journal}{Phys. Rev. Lett.} \textbf{\bibinfo{volume}{42}},
  \bibinfo{pages}{940} (\bibinfo{year}{1979}).

\bibitem[{\citenamefont{Sivers}(1990)}]{Sivers:1990cc}
\bibinfo{author}{\bibfnamefont{D.~W.} \bibnamefont{Sivers}},
  \bibinfo{journal}{Phys. Rev.} \textbf{\bibinfo{volume}{D41}},
  \bibinfo{pages}{83} (\bibinfo{year}{1990}).

\bibitem[{\citenamefont{Collins}(1993)}]{Collins:1993kk}
\bibinfo{author}{\bibfnamefont{J.~C.} \bibnamefont{Collins}},
  \bibinfo{journal}{Nucl. Phys.} \textbf{\bibinfo{volume}{B396}},
  \bibinfo{pages}{161} (\bibinfo{year}{1993}),
  \eprint[http://arXiv.org/abs]{hep-ph/9208213}.

\bibitem[{\citenamefont{Collins and Qiu}(2007)}]{Collins:2007nk}
\bibinfo{author}{\bibfnamefont{J.}~\bibnamefont{Collins}} \bibnamefont{and}
  \bibinfo{author}{\bibfnamefont{J.-W.} \bibnamefont{Qiu}},
  \bibinfo{journal}{Phys. Rev. D} \textbf{\bibinfo{volume}{75}},
  \bibinfo{pages}{114014} (\bibinfo{year}{2007}), \eprint{arXiv:0705.2141
  [hep-ph]}.

\bibitem[{\citenamefont{Ji et~al.}(2005)\citenamefont{Ji, Ma, and
  Yuan}}]{Ji:2004wu}
\bibinfo{author}{\bibfnamefont{X.-d.} \bibnamefont{Ji}},
  \bibinfo{author}{\bibfnamefont{J.-p.} \bibnamefont{Ma}}, \bibnamefont{and}
  \bibinfo{author}{\bibfnamefont{F.}~\bibnamefont{Yuan}},
  \bibinfo{journal}{Phys. Rev.} \textbf{\bibinfo{volume}{D71}},
  \bibinfo{pages}{034005} (\bibinfo{year}{2005}), \eprint{hep-ph/0404183}.

\bibitem[{\citenamefont{Collins and Metz}(2004)}]{Collins:2004nx}
\bibinfo{author}{\bibfnamefont{J.~C.} \bibnamefont{Collins}} \bibnamefont{and}
  \bibinfo{author}{\bibfnamefont{A.}~\bibnamefont{Metz}},
  \bibinfo{journal}{Phys. Rev. Lett.} \textbf{\bibinfo{volume}{93}},
  \bibinfo{pages}{252001} (\bibinfo{year}{2004}), \eprint{hep-ph/0408249}.

\bibitem[{\citenamefont{Boer and Mulders}(1998)}]{Boer:1998nt}
\bibinfo{author}{\bibfnamefont{D.}~\bibnamefont{Boer}} \bibnamefont{and}
  \bibinfo{author}{\bibfnamefont{P.~J.} \bibnamefont{Mulders}},
  \bibinfo{journal}{Phys. Rev.} \textbf{\bibinfo{volume}{D57}},
  \bibinfo{pages}{5780} (\bibinfo{year}{1998}),
  \eprint[http://arXiv.org/abs]{hep-ph/9711485}.

\bibitem[{\citenamefont{Boer}(1999)}]{Boer:1999mm}
\bibinfo{author}{\bibfnamefont{D.}~\bibnamefont{Boer}}, \bibinfo{journal}{Phys.
  Rev.} \textbf{\bibinfo{volume}{D60}}, \bibinfo{pages}{014012}
  (\bibinfo{year}{1999}), \eprint{hep-ph/9902255}.

\bibitem[{\citenamefont{Bacchetta
  et~al.}(2004{\natexlab{a}})\citenamefont{Bacchetta, D'Alesio, Diehl, and
  Miller}}]{Bacchetta:2004jz}
\bibinfo{author}{\bibfnamefont{A.}~\bibnamefont{Bacchetta}},
  \bibinfo{author}{\bibfnamefont{U.}~\bibnamefont{D'Alesio}},
  \bibinfo{author}{\bibfnamefont{M.}~\bibnamefont{Diehl}}, \bibnamefont{and}
  \bibinfo{author}{\bibfnamefont{C.~A.} \bibnamefont{Miller}},
  \bibinfo{journal}{Phys. Rev.} \textbf{\bibinfo{volume}{D70}},
  \bibinfo{pages}{117504} (\bibinfo{year}{2004}{\natexlab{a}}),
  \eprint{hep-ph/0410050}.

\bibitem[{\citenamefont{Bianconi and
  Radici}(2005{\natexlab{a}})}]{Bianconi:2004wu}
\bibinfo{author}{\bibfnamefont{A.}~\bibnamefont{Bianconi}} \bibnamefont{and}
  \bibinfo{author}{\bibfnamefont{M.}~\bibnamefont{Radici}},
  \bibinfo{journal}{Phys. Rev.} \textbf{\bibinfo{volume}{D71}},
  \bibinfo{pages}{074014} (\bibinfo{year}{2005}{\natexlab{a}}),
  \eprint{hep-ph/0412368}.

\bibitem[{\citenamefont{Bianconi and
  Radici}(2005{\natexlab{b}})}]{Bianconi:2005px}
\bibinfo{author}{\bibfnamefont{A.}~\bibnamefont{Bianconi}} \bibnamefont{and}
  \bibinfo{author}{\bibfnamefont{M.}~\bibnamefont{Radici}},
  \bibinfo{journal}{J. Phys.} \textbf{\bibinfo{volume}{G31}},
  \bibinfo{pages}{645} (\bibinfo{year}{2005}{\natexlab{b}}),
  \eprint{hep-ph/0501055}.

\bibitem[{\citenamefont{Burkardt and Hwang}(2004)}]{Burkardt:2003je}
\bibinfo{author}{\bibfnamefont{M.}~\bibnamefont{Burkardt}} \bibnamefont{and}
  \bibinfo{author}{\bibfnamefont{D.~S.} \bibnamefont{Hwang}},
  \bibinfo{journal}{Phys. Rev.} \textbf{\bibinfo{volume}{D69}},
  \bibinfo{pages}{074032} (\bibinfo{year}{2004}), \eprint{hep-ph/0309072}.

\bibitem[{\citenamefont{Collins}(2002)}]{Collins:2002kn}
\bibinfo{author}{\bibfnamefont{J.~C.} \bibnamefont{Collins}},
  \bibinfo{journal}{Phys. Lett.} \textbf{\bibinfo{volume}{B536}},
  \bibinfo{pages}{43} (\bibinfo{year}{2002}), \eprint{hep-ph/0204004}.

\bibitem[{\citenamefont{Aoki et~al.}(1997)\citenamefont{Aoki, Doui, Hatsuda,
  and Kuramashi}}]{Aoki:1996pi}
\bibinfo{author}{\bibfnamefont{S.}~\bibnamefont{Aoki}},
  \bibinfo{author}{\bibfnamefont{M.}~\bibnamefont{Doui}},
  \bibinfo{author}{\bibfnamefont{T.}~\bibnamefont{Hatsuda}}, \bibnamefont{and}
  \bibinfo{author}{\bibfnamefont{Y.}~\bibnamefont{Kuramashi}},
  \bibinfo{journal}{Phys. Rev.} \textbf{\bibinfo{volume}{D56}},
  \bibinfo{pages}{433} (\bibinfo{year}{1997}), \eprint{hep-lat/9608115}.

\bibitem[{\citenamefont{Gockeler et~al.}(2006)}]{Gockeler:2006zu}
\bibinfo{author}{\bibfnamefont{M.}~\bibnamefont{Gockeler}} \bibnamefont{et~al.}
  (\bibinfo{collaboration}{QCDSF}) (\bibinfo{year}{2006}),
  \eprint{hep-lat/0612032}.

\bibitem[{\citenamefont{Bianconi and
  Radici}(2005{\natexlab{c}})}]{Bianconi:2005bd}
\bibinfo{author}{\bibfnamefont{A.}~\bibnamefont{Bianconi}} \bibnamefont{and}
  \bibinfo{author}{\bibfnamefont{M.}~\bibnamefont{Radici}},
  \bibinfo{journal}{Phys. Rev.} \textbf{\bibinfo{volume}{D72}},
  \bibinfo{pages}{074013} (\bibinfo{year}{2005}{\natexlab{c}}),
  \eprint{hep-ph/0504261}.

\bibitem[{\citenamefont{Bianconi and
  Radici}(2006{\natexlab{a}})}]{Bianconi:2005yj}
\bibinfo{author}{\bibfnamefont{A.}~\bibnamefont{Bianconi}} \bibnamefont{and}
  \bibinfo{author}{\bibfnamefont{M.}~\bibnamefont{Radici}},
  \bibinfo{journal}{Phys. Rev.} \textbf{\bibinfo{volume}{D73}},
  \bibinfo{pages}{034018} (\bibinfo{year}{2006}{\natexlab{a}}),
  \eprint{hep-ph/0512091}.

\bibitem[{\citenamefont{Bianconi and
  Radici}(2006{\natexlab{b}})}]{Bianconi:2006hc}
\bibinfo{author}{\bibfnamefont{A.}~\bibnamefont{Bianconi}} \bibnamefont{and}
  \bibinfo{author}{\bibfnamefont{M.}~\bibnamefont{Radici}},
  \bibinfo{journal}{Phys. Rev.} \textbf{\bibinfo{volume}{D73}},
  \bibinfo{pages}{114002} (\bibinfo{year}{2006}{\natexlab{b}}),
  \eprint{hep-ph/0602103}.

\bibitem[{\citenamefont{Collins et~al.}(1994)\citenamefont{Collins, Heppelmann,
  and Ladinsky}}]{Collins:1994kq}
\bibinfo{author}{\bibfnamefont{J.~C.} \bibnamefont{Collins}},
  \bibinfo{author}{\bibfnamefont{S.~F.} \bibnamefont{Heppelmann}},
  \bibnamefont{and} \bibinfo{author}{\bibfnamefont{G.~A.}
  \bibnamefont{Ladinsky}}, \bibinfo{journal}{Nucl. Phys.}
  \textbf{\bibinfo{volume}{B420}}, \bibinfo{pages}{565} (\bibinfo{year}{1994}),
  \eprint[http://arXiv.org/abs]{hep-ph/9305309}.

\bibitem[{\citenamefont{Jaffe et~al.}(1998)\citenamefont{Jaffe, Jin, and
  Tang}}]{Jaffe:1998hf}
\bibinfo{author}{\bibfnamefont{R.~L.} \bibnamefont{Jaffe}},
  \bibinfo{author}{\bibfnamefont{X.}~\bibnamefont{Jin}}, \bibnamefont{and}
  \bibinfo{author}{\bibfnamefont{J.}~\bibnamefont{Tang}},
  \bibinfo{journal}{Phys. Rev. Lett.} \textbf{\bibinfo{volume}{80}},
  \bibinfo{pages}{1166} (\bibinfo{year}{1998}),
  \eprint[http://arXiv.org/abs]{hep-ph/9709322}.

\bibitem[{\citenamefont{Radici et~al.}(2002)\citenamefont{Radici, Jakob, and
  Bianconi}}]{Radici:2001na}
\bibinfo{author}{\bibfnamefont{M.}~\bibnamefont{Radici}},
  \bibinfo{author}{\bibfnamefont{R.}~\bibnamefont{Jakob}}, \bibnamefont{and}
  \bibinfo{author}{\bibfnamefont{A.}~\bibnamefont{Bianconi}},
  \bibinfo{journal}{Phys. Rev.} \textbf{\bibinfo{volume}{D65}},
  \bibinfo{pages}{074031} (\bibinfo{year}{2002}),
  \eprint[http://arXiv.org/abs]{hep-ph/0110252}.

\bibitem[{\citenamefont{van~der Nat}(2005)}]{vanderNat:2005yf}
\bibinfo{author}{\bibfnamefont{P.~B.} \bibnamefont{van~der Nat}}
  (\bibinfo{collaboration}{HERMES}) (\bibinfo{year}{2005}),
  \eprint{hep-ex/0512019}.

\bibitem[{\citenamefont{Joosten}(2005)}]{Joosten:2005}
\bibinfo{author}{\bibfnamefont{R.}~\bibnamefont{Joosten}}
  (\bibinfo{collaboration}{HERMES}) (\bibinfo{year}{2005}),
  \bibinfo{note}{proceedings of the 13th International Workshop on Deep
  Inelastic Scattering (DIS 2005), Madison, Wisconsin, U.S.A., 27 Apr - 1 May
  2005.}

\bibitem[{\citenamefont{Abe et~al.}(2006)}]{Abe:2005zx}
\bibinfo{author}{\bibfnamefont{K.}~\bibnamefont{Abe}} \bibnamefont{et~al.}
  (\bibinfo{collaboration}{Belle}), \bibinfo{journal}{Phys. Rev. Lett.}
  \textbf{\bibinfo{volume}{96}}, \bibinfo{pages}{232002}
  (\bibinfo{year}{2006}), \eprint{hep-ex/0507063}.

\bibitem[{\citenamefont{Bacchetta and Radici}(2006)}]{Bacchetta:2006un}
\bibinfo{author}{\bibfnamefont{A.}~\bibnamefont{Bacchetta}} \bibnamefont{and}
  \bibinfo{author}{\bibfnamefont{M.}~\bibnamefont{Radici}},
  \bibinfo{journal}{Phys. Rev.} \textbf{\bibinfo{volume}{D74}},
  \bibinfo{pages}{114007} (\bibinfo{year}{2006}), \eprint{hep-ph/0608037}.

\bibitem[{\citenamefont{Ceccopieri et~al.}(2007)\citenamefont{Ceccopieri,
  Radici, and Bacchetta}}]{Ceccopieri:2007ip}
\bibinfo{author}{\bibfnamefont{F.~A.} \bibnamefont{Ceccopieri}},
  \bibinfo{author}{\bibfnamefont{M.}~\bibnamefont{Radici}}, \bibnamefont{and}
  \bibinfo{author}{\bibfnamefont{A.}~\bibnamefont{Bacchetta}},
  \bibinfo{journal}{Phys. Lett.} \textbf{\bibinfo{volume}{B650}},
  \bibinfo{pages}{81} (\bibinfo{year}{2007}), \eprint{hep-ph/0703265}.

\bibitem[{\citenamefont{Bacchetta and Radici}(2004)}]{Bacchetta:2004it}
\bibinfo{author}{\bibfnamefont{A.}~\bibnamefont{Bacchetta}} \bibnamefont{and}
  \bibinfo{author}{\bibfnamefont{M.}~\bibnamefont{Radici}},
  \bibinfo{journal}{Phys. Rev.} \textbf{\bibinfo{volume}{D70}},
  \bibinfo{pages}{094032} (\bibinfo{year}{2004}), \eprint{hep-ph/0409174}.

\bibitem[{\citenamefont{Collins et~al.}(1985)\citenamefont{Collins, Soper, and
  Sterman}}]{Collins:1984kg}
\bibinfo{author}{\bibfnamefont{J.~C.} \bibnamefont{Collins}},
  \bibinfo{author}{\bibfnamefont{D.~E.} \bibnamefont{Soper}}, \bibnamefont{and}
  \bibinfo{author}{\bibfnamefont{G.}~\bibnamefont{Sterman}},
  \bibinfo{journal}{Nucl. Phys.} \textbf{\bibinfo{volume}{B250}},
  \bibinfo{pages}{199} (\bibinfo{year}{1985}), \bibinfo{note}{see also A.
  Efremov and A. Radyushkin, Theor. Math. Phys. {\bf 44}, 664 (1981) [ Teor.
  Mat. Fiz. {\bf 44}, 157 (1980)]}.

\bibitem[{\citenamefont{Collins and Soper}(1977)}]{Collins:1977iv}
\bibinfo{author}{\bibfnamefont{J.~C.} \bibnamefont{Collins}} \bibnamefont{and}
  \bibinfo{author}{\bibfnamefont{D.~E.} \bibnamefont{Soper}},
  \bibinfo{journal}{Phys. Rev.} \textbf{\bibinfo{volume}{D16}},
  \bibinfo{pages}{2219} (\bibinfo{year}{1977}).

\bibitem[{\citenamefont{Bomhof et~al.}(2004)\citenamefont{Bomhof, Mulders, and
  Pijlman}}]{Bomhof:2004aw}
\bibinfo{author}{\bibfnamefont{C.~J.} \bibnamefont{Bomhof}},
  \bibinfo{author}{\bibfnamefont{P.~J.} \bibnamefont{Mulders}},
  \bibnamefont{and} \bibinfo{author}{\bibfnamefont{F.}~\bibnamefont{Pijlman}},
  \bibinfo{journal}{Phys. Lett.} \textbf{\bibinfo{volume}{B596}},
  \bibinfo{pages}{277} (\bibinfo{year}{2004}), \eprint{hep-ph/0406099}.

\bibitem[{\citenamefont{Altarelli et~al.}(1979)\citenamefont{Altarelli, Ellis,
  and Martinelli}}]{Altarelli:1979ub}
\bibinfo{author}{\bibfnamefont{G.}~\bibnamefont{Altarelli}},
  \bibinfo{author}{\bibfnamefont{R.~K.} \bibnamefont{Ellis}}, \bibnamefont{and}
  \bibinfo{author}{\bibfnamefont{G.}~\bibnamefont{Martinelli}},
  \bibinfo{journal}{Nucl. Phys.} \textbf{\bibinfo{volume}{B157}},
  \bibinfo{pages}{461} (\bibinfo{year}{1979}).

\bibitem[{\citenamefont{Martin et~al.}(1998)\citenamefont{Martin, Schafer,
  Stratmann, and Vogelsang}}]{Martin:1998rz}
\bibinfo{author}{\bibfnamefont{O.}~\bibnamefont{Martin}},
  \bibinfo{author}{\bibfnamefont{A.}~\bibnamefont{Schafer}},
  \bibinfo{author}{\bibfnamefont{M.}~\bibnamefont{Stratmann}},
  \bibnamefont{and}
  \bibinfo{author}{\bibfnamefont{W.}~\bibnamefont{Vogelsang}},
  \bibinfo{journal}{Phys. Rev.} \textbf{\bibinfo{volume}{D57}},
  \bibinfo{pages}{3084} (\bibinfo{year}{1998}), \eprint{hep-ph/9710300}.

\bibitem[{\citenamefont{Kawamura et~al.}(2007)\citenamefont{Kawamura, Kodaira,
  and Tanaka}}]{Kawamura:2007ze}
\bibinfo{author}{\bibfnamefont{H.}~\bibnamefont{Kawamura}},
  \bibinfo{author}{\bibfnamefont{J.}~\bibnamefont{Kodaira}}, \bibnamefont{and}
  \bibinfo{author}{\bibfnamefont{K.}~\bibnamefont{Tanaka}}
  (\bibinfo{year}{2007}), \eprint{hep-ph/0703079}.

\bibitem[{\citenamefont{Jakob et~al.}(1997)\citenamefont{Jakob, Mulders, and
  Rodrigues}}]{Jakob:1997wg}
\bibinfo{author}{\bibfnamefont{R.}~\bibnamefont{Jakob}},
  \bibinfo{author}{\bibfnamefont{P.~J.} \bibnamefont{Mulders}},
  \bibnamefont{and}
  \bibinfo{author}{\bibfnamefont{J.}~\bibnamefont{Rodrigues}},
  \bibinfo{journal}{Nucl. Phys.} \textbf{\bibinfo{volume}{A626}},
  \bibinfo{pages}{937} (\bibinfo{year}{1997}),
  \eprint[http://arXiv.org/abs]{hep-ph/9704335}.

\bibitem[{\citenamefont{Brodsky et~al.}(2002)\citenamefont{Brodsky, Hwang, and
  Schmidt}}]{Brodsky:2002rv}
\bibinfo{author}{\bibfnamefont{S.~J.} \bibnamefont{Brodsky}},
  \bibinfo{author}{\bibfnamefont{D.~S.} \bibnamefont{Hwang}}, \bibnamefont{and}
  \bibinfo{author}{\bibfnamefont{I.}~\bibnamefont{Schmidt}},
  \bibinfo{journal}{Nucl. Phys.} \textbf{\bibinfo{volume}{B642}},
  \bibinfo{pages}{344} (\bibinfo{year}{2002}), \eprint{hep-ph/0206259}.

\bibitem[{\citenamefont{Gamberg et~al.}(2007)\citenamefont{Gamberg, Goldstein,
  and Schlegel}}]{Gamberg:2007??}
\bibinfo{author}{\bibfnamefont{L.~P.} \bibnamefont{Gamberg}},
  \bibinfo{author}{\bibfnamefont{G.~R.} \bibnamefont{Goldstein}},
  \bibnamefont{and} \bibinfo{author}{\bibfnamefont{M.}~\bibnamefont{Schlegel}}
  (\bibinfo{year}{2007}), \eprint{private communication}.

\bibitem[{\citenamefont{Bacchetta
  et~al.}(2004{\natexlab{b}})\citenamefont{Bacchetta, Schaefer, and
  Yang}}]{Bacchetta:2003rz}
\bibinfo{author}{\bibfnamefont{A.}~\bibnamefont{Bacchetta}},
  \bibinfo{author}{\bibfnamefont{A.}~\bibnamefont{Schaefer}}, \bibnamefont{and}
  \bibinfo{author}{\bibfnamefont{J.-J.} \bibnamefont{Yang}},
  \bibinfo{journal}{Phys. Lett.} \textbf{\bibinfo{volume}{B578}},
  \bibinfo{pages}{109} (\bibinfo{year}{2004}{\natexlab{b}}),
  \eprint{hep-ph/0309246}.

\bibitem[{\citenamefont{Bacchetta
  et~al.}(2007{\natexlab{a}})\citenamefont{Bacchetta, Conti, and
  Radici}}]{futuro}
\bibinfo{author}{\bibfnamefont{A.}~\bibnamefont{Bacchetta}},
  \bibinfo{author}{\bibfnamefont{F.}~\bibnamefont{Conti}}, \bibnamefont{and}
  \bibinfo{author}{\bibfnamefont{M.}~\bibnamefont{Radici}}
  (\bibinfo{year}{2007}{\natexlab{a}}), \eprint{in preparation}.

\bibitem[{\citenamefont{Miyama and Kumano}(1996)}]{Miyama:1995bd}
\bibinfo{author}{\bibfnamefont{M.}~\bibnamefont{Miyama}} \bibnamefont{and}
  \bibinfo{author}{\bibfnamefont{S.}~\bibnamefont{Kumano}},
  \bibinfo{journal}{Comput. Phys. Commun.} \textbf{\bibinfo{volume}{94}},
  \bibinfo{pages}{185} (\bibinfo{year}{1996}), \eprint{hep-ph/9508246}.

\bibitem[{\citenamefont{Hirai et~al.}(1998)\citenamefont{Hirai, Kumano, and
  Miyama}}]{Hirai:1997mm}
\bibinfo{author}{\bibfnamefont{M.}~\bibnamefont{Hirai}},
  \bibinfo{author}{\bibfnamefont{S.}~\bibnamefont{Kumano}}, \bibnamefont{and}
  \bibinfo{author}{\bibfnamefont{M.}~\bibnamefont{Miyama}},
  \bibinfo{journal}{Comput. Phys. Commun.} \textbf{\bibinfo{volume}{111}},
  \bibinfo{pages}{150} (\bibinfo{year}{1998}), \eprint{hep-ph/9712410}.

\bibitem[{\citenamefont{Vogelsang and Yuan}(2005)}]{Vogelsang:2005cs}
\bibinfo{author}{\bibfnamefont{W.}~\bibnamefont{Vogelsang}} \bibnamefont{and}
  \bibinfo{author}{\bibfnamefont{F.}~\bibnamefont{Yuan}},
  \bibinfo{journal}{Phys. Rev.} \textbf{\bibinfo{volume}{D72}},
  \bibinfo{pages}{054028} (\bibinfo{year}{2005}), \eprint{hep-ph/0507266}.

\bibitem[{\citenamefont{Adler et~al.}(2005)}]{Adler:2005in}
\bibinfo{author}{\bibfnamefont{S.~S.} \bibnamefont{Adler}} \bibnamefont{et~al.}
  (\bibinfo{collaboration}{PHENIX}), \bibinfo{journal}{Phys. Rev. Lett.}
  \textbf{\bibinfo{volume}{95}}, \bibinfo{pages}{202001}
  (\bibinfo{year}{2005}), \eprint{hep-ex/0507073}.

\bibitem[{\citenamefont{Bianconi and Radici}(2007)}]{Bianconi:2006mf}
\bibinfo{author}{\bibfnamefont{A.}~\bibnamefont{Bianconi}} \bibnamefont{and}
  \bibinfo{author}{\bibfnamefont{M.}~\bibnamefont{Radici}},
  \bibinfo{journal}{J. Phys. G} \textbf{\bibinfo{volume}{34}},
  \bibinfo{pages}{1595} (\bibinfo{year}{2007}), \eprint{hep-ph/0610317}.

\bibitem[{\citenamefont{Sissakian et~al.}(2005)\citenamefont{Sissakian,
  Shevchenko, Nagaytsev, Denisov, and Ivanov}}]{Sissakian:2005yp}
\bibinfo{author}{\bibfnamefont{A.}~\bibnamefont{Sissakian}},
  \bibinfo{author}{\bibfnamefont{O.}~\bibnamefont{Shevchenko}},
  \bibinfo{author}{\bibfnamefont{A.}~\bibnamefont{Nagaytsev}},
  \bibinfo{author}{\bibfnamefont{O.}~\bibnamefont{Denisov}}, \bibnamefont{and}
  \bibinfo{author}{\bibfnamefont{O.}~\bibnamefont{Ivanov}}
  (\bibinfo{year}{2005}), \eprint{hep-ph/0512095}.

\bibitem[{\citenamefont{Bacchetta
  et~al.}(2007{\natexlab{b}})}]{Bacchetta:2006tn}
\bibinfo{author}{\bibfnamefont{A.}~\bibnamefont{Bacchetta}}
  \bibnamefont{et~al.}, \bibinfo{journal}{JHEP} \textbf{\bibinfo{volume}{02}},
  \bibinfo{pages}{093} (\bibinfo{year}{2007}{\natexlab{b}}),
  \eprint{hep-ph/0611265}.

\bibitem[{\citenamefont{Bianconi et~al.}(2000)\citenamefont{Bianconi, Boffi,
  Jakob, and Radici}}]{Bianconi:1999cd}
\bibinfo{author}{\bibfnamefont{A.}~\bibnamefont{Bianconi}},
  \bibinfo{author}{\bibfnamefont{S.}~\bibnamefont{Boffi}},
  \bibinfo{author}{\bibfnamefont{R.}~\bibnamefont{Jakob}}, \bibnamefont{and}
  \bibinfo{author}{\bibfnamefont{M.}~\bibnamefont{Radici}},
  \bibinfo{journal}{Phys. Rev.} \textbf{\bibinfo{volume}{D62}},
  \bibinfo{pages}{034008} (\bibinfo{year}{2000}),
  \eprint[http://arXiv.org/abs]{hep-ph/9907475}.

\bibitem[{\citenamefont{Boer et~al.}(2003)\citenamefont{Boer, Jakob, and
  Radici}}]{Boer:2003ya}
\bibinfo{author}{\bibfnamefont{D.}~\bibnamefont{Boer}},
  \bibinfo{author}{\bibfnamefont{R.}~\bibnamefont{Jakob}}, \bibnamefont{and}
  \bibinfo{author}{\bibfnamefont{M.}~\bibnamefont{Radici}},
  \bibinfo{journal}{Phys. Rev.} \textbf{\bibinfo{volume}{D67}},
  \bibinfo{pages}{094003} (\bibinfo{year}{2003}), \eprint{hep-ph/0302232}.

\bibitem[{\citenamefont{Bacchetta and Radici}(2003)}]{Bacchetta:2002ux}
\bibinfo{author}{\bibfnamefont{A.}~\bibnamefont{Bacchetta}} \bibnamefont{and}
  \bibinfo{author}{\bibfnamefont{M.}~\bibnamefont{Radici}},
  \bibinfo{journal}{Phys. Rev.} \textbf{\bibinfo{volume}{D67}},
  \bibinfo{pages}{094002} (\bibinfo{year}{2003}), \eprint{hep-ph/0212300}.

\bibitem[{\citenamefont{Gastmans and Wu}(1990)}]{Gastmans:1990xh}
\bibinfo{author}{\bibfnamefont{R.}~\bibnamefont{Gastmans}} \bibnamefont{and}
  \bibinfo{author}{\bibfnamefont{T.~T.} \bibnamefont{Wu}},
  \emph{\bibinfo{title}{The Ubiquitous photon: Helicity method for QED and
  QCD}}, International series of monographs on physics
  (\bibinfo{publisher}{Clarendon}, \bibinfo{address}{Oxford, UK},
  \bibinfo{year}{1990}).

\bibitem[{\citenamefont{Jaffe and Ji}(1992)}]{Jaffe:1992ra}
\bibinfo{author}{\bibfnamefont{R.~L.} \bibnamefont{Jaffe}} \bibnamefont{and}
  \bibinfo{author}{\bibfnamefont{X.}~\bibnamefont{Ji}}, \bibinfo{journal}{Nucl.
  Phys.} \textbf{\bibinfo{volume}{B375}}, \bibinfo{pages}{527}
  (\bibinfo{year}{1992}).

\bibitem[{\citenamefont{Bacchetta et~al.}(2000)\citenamefont{Bacchetta,
  Boglione, Henneman, and Mulders}}]{Bacchetta:1999kz}
\bibinfo{author}{\bibfnamefont{A.}~\bibnamefont{Bacchetta}},
  \bibinfo{author}{\bibfnamefont{M.}~\bibnamefont{Boglione}},
  \bibinfo{author}{\bibfnamefont{A.}~\bibnamefont{Henneman}}, \bibnamefont{and}
  \bibinfo{author}{\bibfnamefont{P.~J.} \bibnamefont{Mulders}},
  \bibinfo{journal}{Phys. Rev. Lett.} \textbf{\bibinfo{volume}{85}},
  \bibinfo{pages}{712} (\bibinfo{year}{2000}),
  \eprint[http://arXiv.org/abs]{hep-ph/9912490}.

\bibitem[{\citenamefont{Jaffe and Saito}(1996)}]{Jaffe:1996ik}
\bibinfo{author}{\bibfnamefont{R.~L.} \bibnamefont{Jaffe}} \bibnamefont{and}
  \bibinfo{author}{\bibfnamefont{N.}~\bibnamefont{Saito}},
  \bibinfo{journal}{Phys. Lett.} \textbf{\bibinfo{volume}{B382}},
  \bibinfo{pages}{165} (\bibinfo{year}{1996}), \eprint{hep-ph/9604220}.

\bibitem[{\citenamefont{Mulders and Rodrigues}(2001)}]{Mulders:2000sh}
\bibinfo{author}{\bibfnamefont{P.~J.} \bibnamefont{Mulders}} \bibnamefont{and}
  \bibinfo{author}{\bibfnamefont{J.}~\bibnamefont{Rodrigues}},
  \bibinfo{journal}{Phys. Rev.} \textbf{\bibinfo{volume}{D63}},
  \bibinfo{pages}{094021} (\bibinfo{year}{2001}), \eprint{hep-ph/0009343}.

\bibitem[{\citenamefont{Jaffe and Manohar}(1989)}]{Jaffe:1989xy}
\bibinfo{author}{\bibfnamefont{R.~L.} \bibnamefont{Jaffe}} \bibnamefont{and}
  \bibinfo{author}{\bibfnamefont{A.}~\bibnamefont{Manohar}},
  \bibinfo{journal}{Phys. Lett.} \textbf{\bibinfo{volume}{B223}},
  \bibinfo{pages}{218} (\bibinfo{year}{1989}).

\bibitem[{\citenamefont{Grosse~Perdekamp}(2002)}]{GrossePerdekamp:2002eb}
\bibinfo{author}{\bibfnamefont{M.}~\bibnamefont{Grosse~Perdekamp}},
  \bibinfo{journal}{Nucl. Phys. Proc. Suppl.} \textbf{\bibinfo{volume}{105}},
  \bibinfo{pages}{71} (\bibinfo{year}{2002}).

\end{thebibliography}

\end{document}